\documentclass[qst,onecolumn,superscriptaddress,nofootinbib]{revtex4}

\usepackage{graphicx}
\usepackage{dcolumn}
\usepackage{bm}
\usepackage{amssymb}
\usepackage{amsfonts}
\usepackage{latexsym}
\usepackage{enumerate}
\usepackage[dvipdfmx]{color} 
\usepackage{amsmath}
\usepackage[dvipdfmx]{epsfig}
\usepackage{times}
\usepackage{algorithm}
\usepackage{algorithmic}
\usepackage{cases}

\newcommand{\red}{\color{black}}
\newtheorem{theorem}{Theorem}

\begin{document}
\widetext
\title{Quantum algorithm for unstructured search of ranked targets}
\author{Kota Tani}
\affiliation{Department of Physics, Chuo University, 1-13-27 Kasuga, Bunkyo-ku, Tokyo 112-8551, Japan}
\author{Shunji Tsuchiya}
\email{tshunji001c@g.chuo-u.ac.jp}
\affiliation{Department of Physics, Chuo University, 1-13-27 Kasuga, Bunkyo-ku, Tokyo 112-8551, Japan}
\author{Seiichiro Tani}
\affiliation{Department of Mathematics, Waseda University, 1-6-1 Nishi-Waseda, Shinjuku-ku, Tokyo 169-8050, Japan}
\author{Yuki Takeuchi}
\email{Takeuchi.Yuki@bk.MItsubishiElectric.co.jp}
\thanks{The current affiliation is Information Technology R\&D Center, Mitsubishi Electric Corporation.}
\affiliation{NTT Communication Science Laboratories, NTT Corporation, 3-1 Morinosato Wakamiya, Atsugi, Kanagawa 243-0198, Japan}
\affiliation{NTT Research Center for Theoretical Quantum Information, NTT Corporation, 3-1 Morinosato Wakamiya, Atsugi, Kanagawa 243-0198, Japan}

\begin{abstract}
Grover's quantum algorithm can find a marked item from an unstructured database faster than any classical algorithm, and hence it has been used for several applications such as cryptanalysis and optimization.
When there exist multiple marked items, Grover's algorithm has the property of finding one of them uniformly at random.
To further broaden the application range, it was generalized so that it finds marked items with probabilities according to their priority by encoding the priority into amplitudes applied by Grover's oracle operator.
In this paper, to achieve a similar generalization, we examine a different encoding that incorporates the priority into phases applied by the oracle operator.
We compare the previous and our oracle operators and observe that which one performs better depends on priority parameters.
Since the priority parameters can be considered as the magnitude of the correlated phase error on Grover's oracle operator, the analysis of our oracle operator also reveals the robustness of the original Grover's algorithm against correlated noises.
We further numerically show that the coherence between multiple marked items tends to increase the probability of finding the most prioritized one in Grover's algorithm with our oracle operator.
\end{abstract}

\maketitle
\section{Introduction}
\label{I}
Several quantum algorithms{\red~\cite{M09}, such as the Harrow-Hassidim-Lloyd (HHL) algorithm~\cite{HHL09} and quantum singular value transformation (QSVT)~\cite{MRTC21},} superior to classical ones have been proposed, and some of them were already demonstrated in small scale experiments {\red by using spins~\cite{CVZLL98,JMH98,VSBYSC01}, photons~\cite{WRRSWVAZ05}, and superconducting qubits~\cite{ZSCXLGZXDHWYZLPWLZ17}}.
Among them, Grover's algorithm~\cite{G97} (for details, see Sec.~\ref{II}) is especially attractive due to its versatility.
It has been applied to numerous applications such as collision and claw finding~\cite{BHT98}, machine learning~\cite{WNHT20,DHLT21}, and optimization~\cite{BBW05,DHHM06}.
Grover's algorithm finds a marked item from an unstructured database quadratically faster than any classical algorithm~\cite{G97,G96} and is optimal in the sense that the dependence of its query complexity on the database size and the number of marked items cannot be improved any further~\cite{BBBV97,BBHT99,Z99}.
Here, query complexity is the number of accesses to the database.

To further broaden the application range, Grover's algorithm has been improved and generalized in various directions.
The success probability of the original Grover's algorithm, i.e., the probability of finding a marked item is close to one, but it is strictly less than one except for the special case.
This issue was solved in Refs.~\cite{H00,L01,BHMT02,RJS22,RJS23} by proposing a modified Grover's algorithm that can find a marked item without failure.
A method of reducing the failure probability to any non-zero small value was also devised in Ref.~\cite{BCWZ99}.
Another property of the original Grover's algorithm is that when there exist multiple marked items in the unstructured database, it outputs one of them uniformly at random, which would imply that the original Grover's algorithm cannot take the priority of the multiple marked items into consideration.
Panchi and Shiyong modified Grover's oracle operator so that the marked items are output with probabilities according to their priority~\cite{PS08}.
More specifically, they encoded the priority into amplitudes applied by Grover's oracle operator [see Eq.~(\ref{oraclePS})].
Grover's algorithm with their modified oracle operator was applied to the cluster head selection in wireless sensor networks~\cite{RK23}.

In this paper, we propose another quantum algorithm for unstructured search with priority.
As a difference from the algorithm in Ref.~\cite{PS08}, we use a phase encoding rather than the amplitude encoding.
In the original Grover's algorithm, the phase $-1$ is applied to every marked item by the oracle operator.
On the other hand, our oracle operator applies $-e^{i\pi\epsilon_i}$ to the $i$th marked item, where $-1\le\epsilon_i\le0$ is the priority parameter.
Since when $\epsilon_i=0$ for all $i$, our oracle operator becomes the original Grover's one, Grover's algorithm with our oracle operator can be considered as a generalization of the original Grover's algorithm.
To clarify which of the phase and amplitude encoding is better for taking the priority into account, we numerically compare our oracle operator and that in Ref.~\cite{PS08} under the condition that there exist two marked items.
As a result, we observe that when a marked item is highly prioritized than another one, our oracle operator would be superior to the existing one.
On the other hand, when the two marked items are similarly prioritized, the existing algorithm becomes better than ours (for details, see Sec.~\ref{IIID}).
We further numerically show that the coherence between the two marked items tends to increase the probability of finding the most prioritized one in Grover's algorithm with our oracle operator.

The analysis of our oracle operator is also related to the robustness of the original Grover's algorithm against correlated phase errors.
This is because the absolute value $|\epsilon_i|$ of the priority parameter can be considered as the noise strength on the oracle operator.
Note that we here suppose that $\epsilon_i$ is an unknown parameter introduced by the noise, while we assume that it is decided by the oracle operator in the previous paragraph.
Furthermore, each marked item is basically represented by multiple qubits, and the absolute value of its priority parameter can depend on all the qubits, and hence we call our errors correlated ones.
{\red Although our error model may not be physically natural, it can happen in, e.g., cloud quantum computation. A service provider of (unsecured) cloud quantum computation can know what quantum circuits are delegated and hence can artificially introduce correlated errors to deceive users.}
As with our analysis, several noise effects on Grover's algorithm have been investigated by considering noisy oracle operators.
Remarkably, when the oracle operator does not work with an arbitrarily small constant  probability, the quadratic speed-up of Grover's algorithm is completely canceled~\cite{RS08}.
The similar results were also shown for the depolarizing and dephasing noises~\cite{R23}.
However, this cancellation does not always occur, and a concrete noise model where the quadratic speed-up survives was found~\cite{HMW03}.
Our noise model was also already investigated under the restriction that there is only a single marked item~\cite{LLZT20,SBW03}, that the noise strength is the same for all marked items~\cite{LSZN99,LL07,TDNTK08,TD24}, or that the noise strength is randomly chosen from $\{-1,0\}$~\cite{ABNR12,KNR18}.
Particularly, Refs.~\cite{TDNTK08,TD24} consider a realistic situation where the noise strength changes every time the oracle operator is applied.
As another noise model, a systematic noise that causes coherent errors was also investigated~\cite{DZKK24}.
Our results, together with these existing results, would deepen the understanding on the noise robustness of Grover's algorithm.

As another related work, a quantum algorithm for unstructured search of ranked targets was also proposed in Ref.~\cite{SKH22}.
This algorithm uses multiple kinds of oracle operators, while ours and the algorithm in Ref.~\cite{PS08} use a single kind of the oracle operators, respectively.
Although we consider the problem of finding ranked targets from an unstructured database, Sun and Wu consider that of finding a single target from a weighted database~\cite{SW24}.
More specifically, they use the original Grover's oracle operator but modify the initial state (and diffusion operator) to represent the weighted database.

\section{Grover's algorithm}
\label{II}
Since our oracle operator is constructed by modifying Grover's oracle operator, we first review the original Grover's algorithm~\cite{G97,LMP03}.
For a given function $f:\mathcal{X}\equiv\{0,1,\ldots,n-1\}\rightarrow\{0,1\}$ with a natural number $n$, its purpose is to find a marked item $x\in\mathcal{X}$ satisfying $f(x)=1$.
It is worth mentioning that such $x$ is, in general, not unique~\cite{GK17}.
Let $m\equiv\sum_{x\in\mathcal{X}}f(x)$ be the number of $x$'s satisfying $f(x)=1$.
For simplicity, we assume $1\le m\ll n$ in this section.
To find a marked item (with a sufficiently high probability), any classical algorithm requires  $\Omega(n/m)$ queries to an oracle for $f$ in the worst case.
On the other hand, Grover's algorithm can do the same thing with only $O(\sqrt{n/m})$ quantum queries.

Grover's algorithm proceeds as follows:
\begin{enumerate}
\item Prepare the initial state
\begin{eqnarray}
\label{Groverini}
|\psi(0)\rangle\equiv\cfrac{1}{\sqrt{n}}\sum_{x=0}^{n-1}|x\rangle.
\end{eqnarray}
Here, we use the notation $|\psi(0)\rangle$ to make Eq.~(\ref{Groverini}) consistent with Eq.~(\ref{Grovert}).
\item Apply the unitary operator $G\equiv DO_f$ to the initial state in Eq.~(\ref{Groverini}) $t$ times, where
\begin{eqnarray}
D\equiv 2|\psi(0)\rangle\langle\psi(0)|-I^{(n)}
\end{eqnarray}
is Grover's diffusion operator, $I^{(d)}$ is the $d$-dimensional identity operator for any natural number $d$, and
\begin{eqnarray}
\label{oracle}
O_f\equiv I^{(n)}-2\sum_{x\in\mathcal{X}_1}|x\rangle\langle x|
\end{eqnarray}
is the oracle operator.
Here, $\mathcal{X}_b\equiv\{x|x\in\mathcal{X},f(x)=b\}$, and hence $\mathcal{X}_0\cup\mathcal{X}_1=\mathcal{X}$, and $\mathcal{X}_0\cap\mathcal{X}_1=\varnothing$.
At the end of this step, the state is
\begin{eqnarray}
\label{Grovert}
|\psi(t)\rangle\equiv G^t|\psi(0)\rangle.
\end{eqnarray}
\item Measure the final state $|\psi(t)\rangle$ in the computational basis $\{|x\rangle\}_{x\in\mathcal{X}}$.
\item Output the measurement outcome $x$ as a solution.
\end{enumerate}

\begin{figure}[t]
\includegraphics[width=8.5cm, clip]{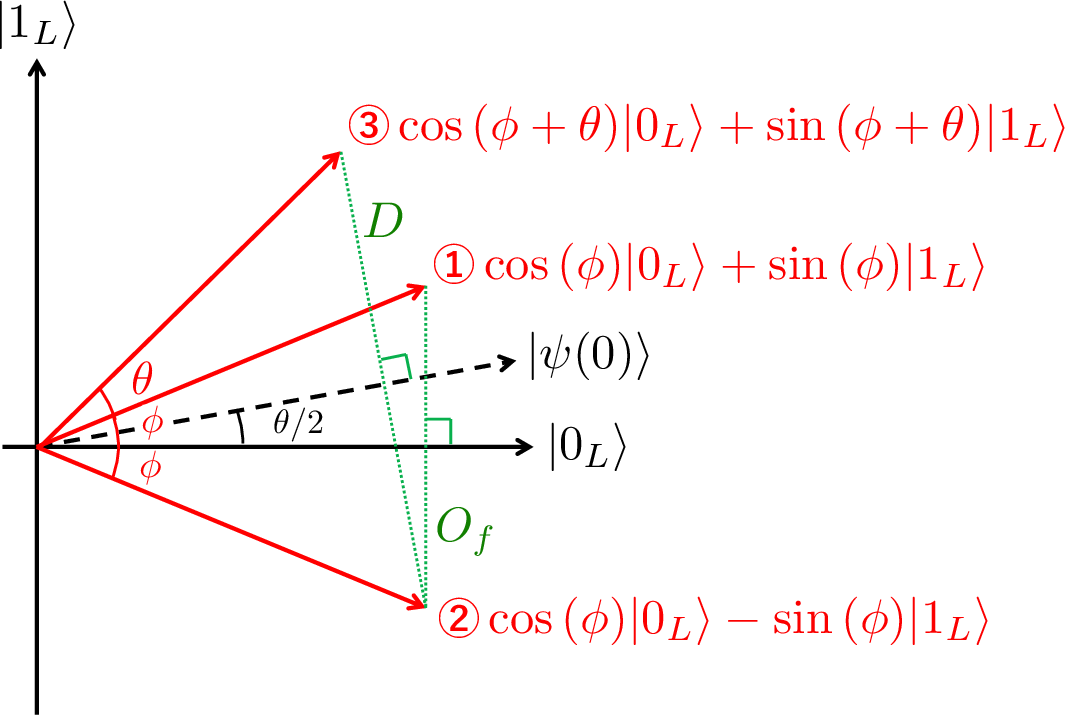}
\caption{Geometric interpretation of $G\equiv DO_f$. The oracle operator $O_f$ transforms $\cos{(\phi)}|0_L\rangle+\sin{(\phi)}|1_L\rangle$ to the quantum state in Eq.~(\ref{stateoracle}) for any real number $\phi$, which can be interpreted as the reflection along the horizontal axis $|0_L\rangle$. Grover's diffusion operator $D$ transforms the quantum state in Eq.~(\ref{stateoracle}) to that in Eq.~(\ref{G}). It is interpreted as the reflection along the dotted axis corresponding to the initial state $|\psi(0)\rangle$. These interpretations imply that $|\psi(0)\rangle$ gets close to the superposition $|1_L\rangle$ of marked items by repeating the implementation of $G$.}
\label{Groverfig}
\end{figure}

The query to the oracle for $f$ is implemented with the oracle operator $O_f$ because it multiplies $(-1)^{f(x)}$ to $|x\rangle$.
Therefore, the query complexity of Grover's algorithm is the number $t$ of uses of $O_f$.
We derive $t=O(\sqrt{n/m})$ by using a geometric interpretation of $G$ given in Fig.~\ref{Groverfig}.
Although Grover's algorithm uses multiple qubits (more precisely, $O(\log{n})$ qubits), its behaviour can be captured as a quantum algorithm running on a single logical qubit with the orthonormal basis
\begin{eqnarray}
\label{logical0}
\left\{|0_L\rangle\equiv\cfrac{1}{\sqrt{n-m}}\sum_{x\in\mathcal{X}_0}|x\rangle,|1_L\rangle\equiv\cfrac{1}{\sqrt{m}}\sum_{x\in\mathcal{X}_1}|x\rangle\right\}.
\end{eqnarray}
By using the basis in Eq.~(\ref{logical0}), the initial state in Eq.~(\ref{Groverini}) is rewritten as
\begin{eqnarray}
|\psi(0)\rangle=\cos{\left(\cfrac{\theta}{2}\right)}|0_L\rangle+\sin{\left(\cfrac{\theta}{2}\right)}|1_L\rangle,
\end{eqnarray}
where $0<\theta<\pi$, and
\begin{eqnarray}
\label{theta}
\theta\equiv2\sin^{-1}{\left(\sqrt{\cfrac{m}{n}}\right)}\simeq2\sqrt{\cfrac{m}{n}}.
\end{eqnarray}
Here, the last approximation comes from the assumption that $m\ll n$.
On the other hand, the direct calculation shows that
\begin{eqnarray}
O_f\left[\cos{(\phi)}|0_L\rangle+\sin{(\phi)}|1_L\rangle\right]=\cos{(\phi)}|0_L\rangle-\sin{(\phi)}|1_L\rangle
\label{stateoracle}
\end{eqnarray}
for any real number $\phi$.
Then, Grover's diffusion operator $D$ transforms the quantum state in Eq.~(\ref{stateoracle}) to
\begin{eqnarray}
\label{G}
DO_f\left[\cos{(\phi)}|0_L\rangle+\sin{(\phi)}|1_L\rangle\right]=\cos{(\phi+\theta)}|0_L\rangle+\sin{(\phi+\theta)}|1_L\rangle.
\end{eqnarray}
The transformations in Eqs.~(\ref{stateoracle}) and (\ref{G}) can be interpreted as the reflections in Fig.~\ref{Groverfig}.
This geometrical interpretation implies that the angle $\theta$ is added every time we apply $G$, and thus the final state of Grover's algorithm is
\begin{eqnarray}
\label{fstate}
|\psi(t)\rangle=\cos{\left(\cfrac{2t+1}{2}\theta\right)}|0_L\rangle+\sin{\left(\cfrac{2t+1}{2}\theta\right)}|1_L\rangle.
\end{eqnarray}
The success probability (i.e., the probability of obtaining $x$ satisfying $f(x)=1$ in step 3) $\sin^2{((2t+1)\theta/2)}$ is maximized when
\begin{eqnarray}
\label{queryGrover}
t=\left\lfloor\cfrac{\pi}{2\theta}\right\rfloor=O\left(\sqrt{\cfrac{n}{m}}\right)
\end{eqnarray}
with $\lfloor\cdot\rfloor$ being the floor function.
Note that in the exceptional case of $\pi/(2\theta)$ being an integer, $t=\lfloor\pi/(2\theta)\rfloor-1$ also maximizes the success probability.

It is worth mentioning that since $|1_L\rangle$ is the equal superposition of $m$ marked items [see Eq.~(\ref{logical0})], Grover's algorithm outputs each correct answer with the same probability
\begin{eqnarray}
\cfrac{\sin^2{\left(\left(\left\lfloor\cfrac{\pi}{2\theta}\right\rfloor+\cfrac{1}{2}\right)\theta\right)}}{m}\simeq\cfrac{1}{m}.
\end{eqnarray}
Our subject in this paper is to bias the probability distribution to reflect the priority of each answer by modifying the oracle operator in Eq.~(\ref{oracle}).

\section{Generalized Grover's algorithm with ranked targets}
In this section, we give our main results.
In Sec.~\ref{IIIA}, we explain the behaviour of Grover's algorithm for our oracle operator and numerically evaluate it.
In Sec.~\ref{IIIB}, we give an analytical evaluation of our oracle operator under a concrete condition.
As a result, we show that its success probability can be calculated by solving a cubic equation.
We also obtain a sufficient condition for that the most prioritized items are more frequently observed than the other items by deriving approximations of success probabilities of Grover's algorithm with our oracle operator.
In Sec.~\ref{IIIC}, we consider the case of $m=2$ and numerically observe that the coherence between the two marked items tends to increase the probability of finding the most prioritized one in Grover's algorithm with our oracle operator.
In this sense, our oracle operator effectively uses the quantum effect.
In Sec.~\ref{IIID}, we compare our oracle operator with that in Ref.~\cite{PS08}.

\subsection{Unstructured search of ranked targets}
\label{IIIA}
As stated in Sec.~\ref{II}, the original Grover's algorithm treats all marked items equally.
However, in general, some of the marked items may be prioritized than the others.
For example, when a given oracle can only decide the likelihood that an input $x$ satisfies $f(x)=1$, the marked items would be ranked depending on their likelihood.
As stated in Sec.~\ref{I}, the same situation arises also when the oracle operator is affected by correlated phase errors.
Our purpose is to devise a quantum algorithm that finds the marked items with probabilities according to their priority given an oracle containing the information of the marked items and their priority parameters.
Note that we assume that the number of kinds of the priority parameters and that of the marked items for each kind of the priority parameters are known.
Therefore, the number $m$ of the marked items is also known.
More formally, let $-1\le\epsilon_x\le 0$ be a priority parameter for any marked item $x\in\mathcal{X}_1$.
When two marked items $x$ and $y$ satisfy $\epsilon_x>\epsilon_y$, we would like to find $x$ with a higher probability than that of $y$.
It is worth mentioning that the values of the priority parameters are not given in advance except for that the priority parameter of the most prioritized marked item(s) is $0$.

To this end, it would be natural to apply the maximum phase shift $-1$ and no phase shift to $|x\rangle$ when the priority parameter takes its maximum (i.e., $\epsilon_x=0$) and minimum (i.e., $\epsilon_x=-1$), respectively.
This functionality is achieved by replacing the oracle operator $O_f$ in Eq.~(\ref{oracle}) with
\begin{eqnarray}
\label{rankedoracle}
\tilde{O}_f(\vec{\epsilon})\equiv I^{(n)}-\sum_{x\in\mathcal{X}_1}(1+e^{i\pi\epsilon_x})|x\rangle\langle x|,
\end{eqnarray}
where $\vec{\epsilon}\equiv{(\epsilon_x)}_{x\in\mathcal{X}_1}$.
We can easily check that
\begin{eqnarray}
\tilde{O}_f(\vec{\epsilon})|x\rangle=-e^{i\pi\epsilon_x}|x\rangle
\end{eqnarray}
for any $x$, and hence $|x\rangle$ with $\epsilon_x=0$ and $\epsilon_x=-1$ satisfies $\tilde{O}_f(\vec{\epsilon})|x\rangle=-|x\rangle$ and $\tilde{O}_f(\vec{\epsilon})|x\rangle=|x\rangle$, respectively.
By definition, it is trivial that $\tilde{O}_f(\vec{\epsilon})$ becomes the original oracle operator $O_f$ when $\epsilon_x=0$ holds for all $x\in\mathcal{X}_1$.
Therefore, we can say that our oracle operator is a generalization of Grover's oracle operator.
On the other hand, when $\epsilon_x=-1$ for all $x$, our oracle operator $\tilde{O}_f(\vec{\epsilon})$ becomes $I^{(n)}$, which corresponds to the case where there is no marked item.
{\red In Appendix~\ref{D}, we consider a graph coloring problem (GCP) as a potential application and explain how to construct our oracle operator for the GCP.} 

Our quantum algorithm is the same as Grover's algorithm except for that the oracle operator is replaced with $\tilde{O}_f(\vec{\epsilon})$, and hence the final state of our algorithm is
\begin{eqnarray}
\label{iniour}
|\tilde{\psi}(\vec{\epsilon},t)\rangle\equiv\left[D\tilde{O}_f(\vec{\epsilon})\right]^t|\psi(0)\rangle.
\end{eqnarray}
Our construction is somewhat similar but not identical to that in Ref.~\cite{L01}.
Long introduced a common phase $\epsilon$ for the oracle and diffusion operators to make the success probability of Grover's algorithm exactly $1$, but we do not change the diffusion operator $D$.
For any $1\le i\le m$, we define $\tilde{x}_i$ and $P_{\rm suc}^{(\tilde{x}_i)}(\vec{\epsilon},t)\equiv|\langle \tilde{x}_i|\tilde{\psi}(\vec{\epsilon},t)\rangle|^2$ as the $i$th marked item and the probability of obtaining $\tilde{x}_i$ in our quantum algorithm, respectively.

\begin{figure}[t]
\includegraphics[width=18cm, clip]{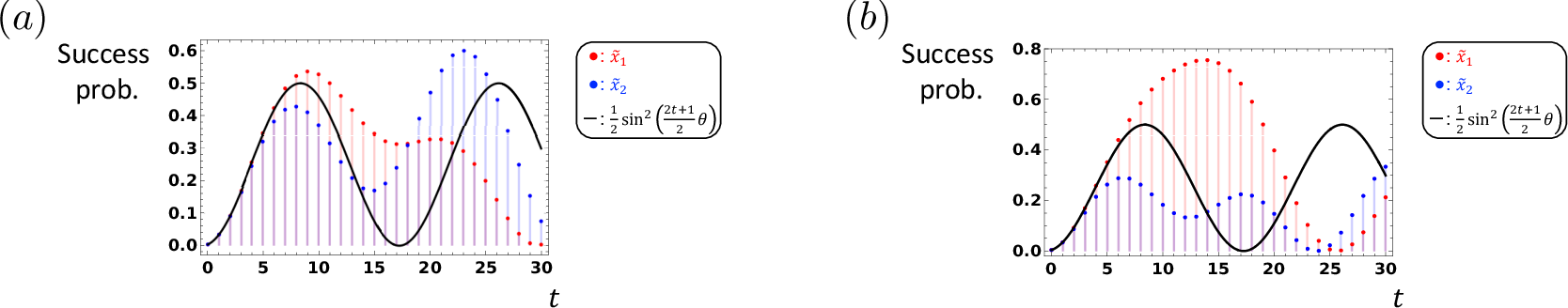}
\caption{Success probabilities of our quantum algorithm. In this figure, we set $n=256$, $m=2$, and $\epsilon_{\tilde{x}_1}=0$. The red and blue dotted curves represent $P_{\rm suc}^{(\tilde{x}_1)}(\vec{\epsilon},t)$ and $P_{\rm suc}^{(\tilde{x}_2)}(\vec{\epsilon},t)$, respectively. The black curves correspond to the probability $\sin^2{((2t+1)\theta/2)}/m$ of the original Grover's algorithm finding a single marked item $\tilde{x}_1$ or $\tilde{x}_2$. (a) The case of $\epsilon_{\tilde{x}_2}=-0.05$. (b) The case of $\epsilon_{\tilde{x}_2}=-0.1$.}
\label{sucprobfigure}
\end{figure}

\begin{figure}[t]
\includegraphics[width=18cm, clip]{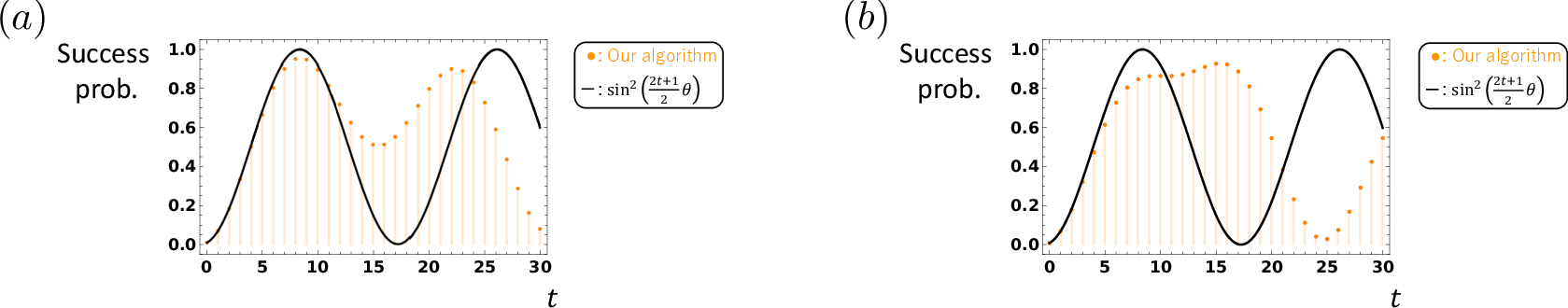}
\caption{Overall success probabilities of our quantum algorithm. In this figure, we set $n=256$, $m=2$, and $\epsilon_{\tilde{x}_1}=0$. The orange dotted curves represent $\sum_{x\in\mathcal{X}_1}P_{\rm suc}^{(x)}(\vec{\epsilon},t)$. The black curves correspond to the overall success probability $\sin^2{((2t+1)\theta/2)}$ of the original Grover's algorithm. (a) The case of $\epsilon_{\tilde{x}_2}=-0.05$. (b) The case of $\epsilon_{\tilde{x}_2}=-0.1$.}
\label{allsucprobfigure}
\end{figure}

To show, in a tangible way, that our quantum algorithm correctly prioritizes marked items, we perform several numerical simulations with $n=256$, $m=2$, and $\epsilon_{\tilde{x}_1}=0$.
To this end, we use Mathematica.
The first numerical simulation is given in Fig.~\ref{sucprobfigure}.
It reveals that the first marked item $\tilde{x}_1$ with the highest priority parameter $\epsilon_{\tilde{x}_1}=0$ is more frequently observed than the second marked item $\tilde{x}_2$ with a lower priority parameter $\epsilon_{\tilde{x}_2}<0$ by choosing an appropriate $t$ such as $\lfloor\pi/(2\theta)\rfloor=8$ [see Eq.~(\ref{queryGrover})].
More concretely, compared with the original Grover's algorithm, the observation of $\tilde{x}_1$ is facilitated, but that of $\tilde{x}_2$ is suppressed.
On the other hand, at inappropriate values of $t$ such as $t=30$, $\tilde{x}_2$ is more frequently observed than $\tilde{x}_1$.
To avoid this unfavourable situation, we will give an analytical sufficient condition on $t$ in Sec.~\ref{IIIB} under the assumption that $|\epsilon_{\tilde{x}_2}|$ is sufficiently small but not $0$.
It is also worth mentioning that the unfavourable situation may disappear by increasing the database size $n$(, which would be expected from Fig.~\ref{SucProbLfigure} in the next subsection).
We then compare the overall success probability $\sum_{x\in\mathcal{X}_1}P_{\rm suc}^{(x)}(\vec{\epsilon},t)$ of our algorithm with that of the original Grover's algorithm in Fig.~\ref{allsucprobfigure}.
From this comparison, we can deduce that the priority is yielded by sacrificing the overall success probability.

\begin{figure}[t]
\includegraphics[width=18cm, clip]{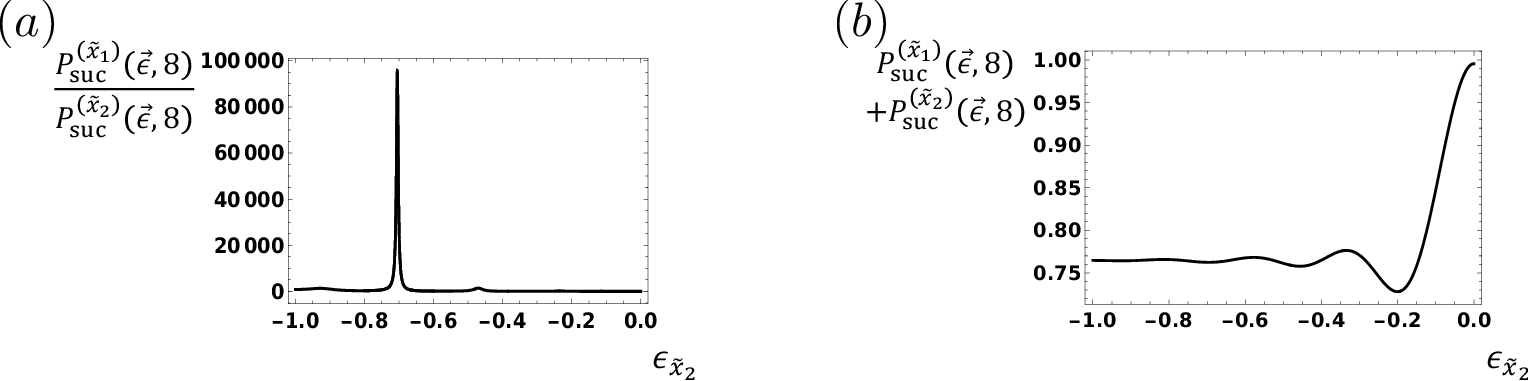}
\caption{Dependence of our quantum algorithm on $\epsilon_{\tilde{x}_2}$. In this figure, we set $n=256$, $m=2$, $\epsilon_{\tilde{x}_1}=0$, and $t=\lfloor\pi/(2\theta)\rfloor=8$. (a) The ratio of $P_{\rm suc}^{(\tilde{x}_1)}(\vec{\epsilon},8)$ to $P_{\rm suc}^{(\tilde{x}_2)}(\vec{\epsilon},8)$. (b) The overall success probability $\sum_{i=1}^2P_{\rm suc}^{(\tilde{x}_i)}(\vec{\epsilon},8)$.}
\label{priorityfigure}
\end{figure}

\begin{figure}[t]
\includegraphics[width=18cm, clip]{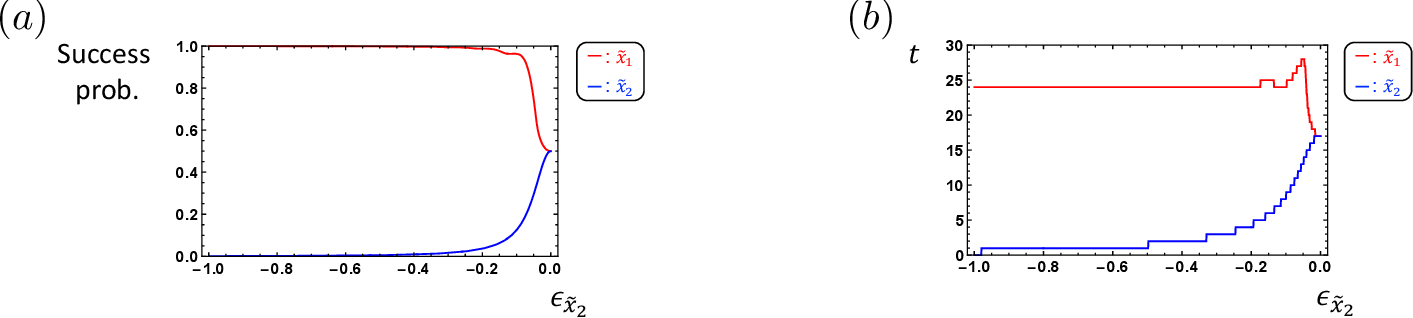}
\caption{Dependence of our quantum algorithm on $\epsilon_{\tilde{x}_2}$. In this figure, we set $n=1000$, $m=2$, and $\epsilon_{\tilde{x}_1}=0$. (a) The first local maximum values of the success probabilities $P_{\rm suc}^{(\tilde{x}_1)}(\vec{\epsilon},t)$ and $P_{\rm suc}^{(\tilde{x}_2)}(\vec{\epsilon},t)$ over $t$. 
The red and blue curves correspond to $\tilde{x}_1$ and $\tilde{x}_2$, respectively.
(b) The numbers $t$ of queries such that $P_{\rm suc}^{(\tilde{x}_1)}(\vec{\epsilon},t)$ and $P_{\rm suc}^{(\tilde{x}_2)}(\vec{\epsilon},t)$ take the first local maxima. The red and blue curves correspond to $\tilde{x}_1$ and $\tilde{x}_2$, respectively.}
\label{interferecenpmaxfigure}
\end{figure}

To evaluate the flexibility of our quantum algorithm, we next calculate the ratio $P_{\rm suc}^{(\tilde{x}_1)}(\vec{\epsilon},8)/P_{\rm suc}^{(\tilde{x}_2)}(\vec{\epsilon},8)$ and the overall success probability $\sum_{i=1}^2P_{\rm suc}^{(\tilde{x}_i)}(\vec{\epsilon},8)$ while varying $\epsilon_{\tilde{x}_2}$ from $-1$ to $0$.
The ratio represents to what extent we can prioritize the two marked items by using our quantum algorithm.
As seen from Fig.~\ref{sucprobfigure}, $P_{\rm suc}^{(\tilde{x}_1)}(\vec{\epsilon},t)$ and $P_{\rm suc}^{(\tilde{x}_2)}(\vec{\epsilon},t)$ take the first local maxima at different $t$, and hence we cannot uniquely determine the optimal value of $t$.
This is why we tentatively adopt the original Grover's query complexity $t=\lfloor\pi/(2\theta)\rfloor=8$.
The result of this calculation is given in Fig.~\ref{priorityfigure}.
From Fig.~\ref{priorityfigure}(a), we can observe that the ratio $P_{\rm suc}^{(\tilde{x}_1)}(\vec{\epsilon},8)/P_{\rm suc}^{(\tilde{x}_2)}(\vec{\epsilon},8)$ becomes $1$ and $95764.3$ when $\epsilon_{\tilde{x}_2}$ is $0$ and $-0.704696$, respectively.
Despite this high flexibility, our algorithm keeps the high overall success probability larger than $0.72$ [see Fig.~\ref{priorityfigure}(b)].
In other words, by considering $|\epsilon_{\tilde{x}_2}|$ as the noise strength, this figure implies that Grover's algorithm is robust against correlated phase errors.
This is because $0.72$ is not significantly smaller than the ideal success probability $\sin^2{(17\theta/2)}\simeq0.998$ of the original Grover's algorithm.

We investigate our algorithm in more detail by increasing the database size to $n=1000$ and clarifying a difference between each marked item.
In Fig.~\ref{interferecenpmaxfigure}(a), we first numerically derive the $\epsilon_{\tilde{x}_2}$-dependence of the first local maximum values of the success probabilities $P_{\rm suc}^{(\tilde{x}_1)}(\vec{\epsilon},t)$ and $P_{\rm suc}^{(\tilde{x}_2)}(\vec{\epsilon},t)$ over $t$.
We notice that the probability of finding the first marked item $\tilde{x}_1$ becomes close to $1$ and almost invariant when $\epsilon_{\tilde{x}_2}\lesssim -0.2$.
This would also be a circumstantial evidence that the original Grover's algorithm is robust against correlated phase errors.
Furthermore, since the point corresponding to $\epsilon_{\tilde{x}_2}=0$ [i.e., the intersection of the red and blue lines in Fig.~\ref{interferecenpmaxfigure}(a)] represents the half of the success probability of the original Grover's algorithm, our algorithm enhances the observation of $\tilde{x}_1$ but suppresses that of $\tilde{x}_2$.
This is consistent with Fig.~\ref{sucprobfigure} because the first local maximum values of the red and blue curves are larger and smaller than those of the black curves, respectively.
Similarly, in Fig.~\ref{interferecenpmaxfigure}(b), we give the $\epsilon_{\tilde{x}_2}$-dependence of the query complexity $t$ such that $P_{\rm suc}^{(\tilde{x}_1)}(\vec{\epsilon},t)$ and $P_{\rm suc}^{(\tilde{x}_2)}(\vec{\epsilon},t)$ take the first local maxima.
We can observe essentially the same behaviour as that of Fig.~\ref{interferecenpmaxfigure}(a).
That is, the query complexity corresponding to $\tilde{x}_1$ (i.e., the red line) becomes invariant when $\epsilon_{\tilde{x}_2}\lesssim -0.2$, and the query complexity corresponding $\tilde{x}_1$ and $\tilde{x}_2$ are increased and decreased compared with that of the original Grover's algorithm [i.e., the intersection of the red and blue lines in Fig.~\ref{interferecenpmaxfigure}(b)], respectively.
This figure also shows that $P_{\rm suc}^{(\tilde{x}_1)}(\vec{\epsilon},t)$ and $P_{\rm suc}^{(\tilde{x}_2)}(\vec{\epsilon},t)$ take the first local maxima at different values of $t$ except when $\epsilon_{\tilde{x}_2}\simeq0$, and hence the success probabilities corresponding to the red and blue lines in Fig.~\ref{interferecenpmaxfigure}(a) are not simultaneously achieved in general.

\subsection{Analytical evaluation}
\label{IIIB}
In this section, we analytically evaluate our quantum algorithm.
As an issue to be solved, the geometrical interpretation shown in Fig.~\ref{Groverfig} does not work because of the modification on the oracle operator in Eq.~(\ref{rankedoracle}).
Therefore, we introduce a different approach to obtain the final state $|\tilde{\psi}(\vec{\epsilon},t)\rangle$ based on the diagonalization of $D\tilde{O}_f(\vec{\epsilon})$.
Although our approach is applicable to the general scenario, for simplicity, we particularly consider the situation that marked items can be divided into two sets $\mathcal{X}_{1,0}$ and $\mathcal{X}_{1,\tilde{\epsilon}}$ such that (i) $\forall x\in\mathcal{X}_{1,0}$, the value of the priority parameter is $\epsilon_x=0$, (ii) $\forall x\in\mathcal{X}_{1,\tilde{\epsilon}}$, the value of the priority parameter is $\epsilon_x=\tilde{\epsilon}$, and (iii) the cardinarities of the two sets are equal, i.e., $|\mathcal{X}_{1,0}|=|\mathcal{X}_{1,\tilde{\epsilon}}|=m/2$.
It is trivial that $\mathcal{X}_{1,0}\cup\mathcal{X}_{1,\tilde{\epsilon}}=\mathcal{X}_1$ and $\mathcal{X}_{1,0}\cap\mathcal{X}_{1,\tilde{\epsilon}}=\varnothing$ hold.
This situation is not general but includes all the situations considered in Sec.~\ref{IIIA}.

In this case, our oracle operator in Eq.~(\ref{rankedoracle}) becomes
\begin{eqnarray}
\label{rankedoracle2}
\tilde{O}_f(\tilde{\epsilon})=I^{(n)}-2\sum_{x\in\mathcal{X}_{1,0}}|x\rangle\langle x|-\left(1+e^{i\pi\tilde{\epsilon}}\right)\sum_{x\in\mathcal{X}_{1,\tilde{\epsilon}}}|x\rangle\langle x|.
\end{eqnarray}
By following the idea in Sec.~\ref{II}, we introduce the three logical basis states as follows:
\begin{eqnarray}
|1'_L\rangle\equiv\sqrt{\cfrac{2}{m}}\sum_{x\in\mathcal{X}_{1,0}}|x\rangle,
\end{eqnarray}
\begin{eqnarray}
|1''_L\rangle\equiv\sqrt{\cfrac{2}{m}}\sum_{x\in\mathcal{X}_{1,\tilde{\epsilon}}}|x\rangle,
\end{eqnarray}
and $|0_L\rangle$ in Eq.~(\ref{logical0}).
The quantum state during the execution of our algorithm can be written as a superposition of these basis states.
In fact, the initial state, Grover's diffusion operator, and our oracle operator are rewritten as
\begin{eqnarray}
\label{idealini}
|\psi(0)\rangle=\cos{\left(\cfrac{\theta}{2}\right)}|0_L\rangle+\sin{\left(\cfrac{\theta}{2}\right)}\cfrac{|1'_L\rangle+|1''_L\rangle}{\sqrt{2}}=\cfrac{1}{\sqrt{2}}\begin{pmatrix}
\sqrt{2}\cos{\left(\cfrac{\theta}{2}\right)} \\
\sin{\left(\cfrac{\theta}{2}\right)} \\
\sin{\left(\cfrac{\theta}{2}\right)} \\
\end{pmatrix},
\end{eqnarray}
\begin{eqnarray}
\label{Dmatrix}
D=\begin{pmatrix}
\cos{\left(\theta\right)} & \cfrac{\sin{\left(\theta\right)}}{\sqrt{2}} & \cfrac{\sin{\left(\theta\right)}}{\sqrt{2}} \\
\cfrac{\sin{\left(\theta\right)}}{\sqrt{2}} & -\cfrac{1+\cos{\left(\theta\right)}}{2} & \cfrac{1-\cos{\left(\theta\right)}}{2} \\
\cfrac{\sin{\left(\theta\right)}}{\sqrt{2}} & \cfrac{1-\cos{\left(\theta\right)}}{2} & -\cfrac{1+\cos{\left(\theta\right)}}{2} \\
\end{pmatrix},
\end{eqnarray}
and
\begin{eqnarray}
\label{Omatrix}
\tilde{O}_f(\tilde{\epsilon})=\begin{pmatrix}
1 & 0 & 0 \\
0 & -1 & 0 \\
0 & 0 & -e^{i\pi\tilde{\epsilon}} \\
\end{pmatrix},
\end{eqnarray}
respectively.
For any $1\le j\le 3$, let $\lambda_j$ and $|\phi_j\rangle$ be the $j$th eigenvalue of $D\tilde{O}_f(\tilde{\epsilon})$ and the normalized eigenvector associated with $\lambda_j$, respectively.
From Eq.~(\ref{iniour}), the final state is
\begin{eqnarray}
\label{finalanalytical}
|\tilde{\psi}(\vec{\epsilon},t)\rangle=\sum_{j=1}^3\lambda_j^t\langle\phi_j|\psi(0)\rangle|\phi_j\rangle,
\end{eqnarray}
and thus it is sufficient to derive $\{|\phi_j\rangle\}_{j=1}^3$ and $\{\lambda_j\}_{j=1}^3$ for our purpose.

From Eqs.~(\ref{Dmatrix}) and (\ref{Omatrix}), we obtain
\begin{eqnarray}
\label{DOmf}
D\tilde{O}_f(\tilde{\epsilon})=\begin{pmatrix}
\cos{\left(\theta\right)} & -\cfrac{\sin{\left(\theta\right)}}{\sqrt{2}} & -e^{i\pi\tilde{\epsilon}}\cfrac{\sin{\left(\theta\right)}}{\sqrt{2}} \\
\cfrac{\sin{\left(\theta\right)}}{\sqrt{2}} & \cfrac{1+\cos{\left(\theta\right)}}{2} & -e^{i\pi\tilde{\epsilon}}\cfrac{1-\cos{\left(\theta\right)}}{2} \\
\cfrac{\sin{\left(\theta\right)}}{\sqrt{2}} & -\cfrac{1-\cos{\left(\theta\right)}}{2} & e^{i\pi\tilde{\epsilon}}\cfrac{1+\cos{\left(\theta\right)}}{2} \\
\end{pmatrix}.
\end{eqnarray}
When $1\le m\le n/2$, the eigenvector of this matrix associated with $\lambda\in\{\lambda_j\}_{j=1}^3$ is
\begin{eqnarray}
\nonumber
\cfrac{1}{\sqrt{2|1-\lambda|^2\sin^2{\left(\theta\right)}+|1+\lambda|^2\left[1-\cos{\left(\theta\right)}\right]^2+|1+\cos{\left(\theta\right)}-\lambda[1+3\cos{\left(\theta\right)}-2\lambda]|^2}}\begin{pmatrix}
\sqrt{2}(1-\lambda)\sin{\left(\theta\right)}e^{i(\pi\tilde{\epsilon}+\phi)} \\
-(1+\lambda)[1-\cos{\left(\theta\right)}]e^{i(\pi\tilde{\epsilon}+\phi)} \\
|1+\cos{\left(\theta\right)}-\lambda[1+3\cos{\left(\theta\right)}-2\lambda]| \\
\end{pmatrix}
,\\
\label{eigenvector}
\end{eqnarray}
where
\begin{eqnarray}
e^{i\phi}\equiv\cfrac{|1+\cos{\left(\theta\right)}-\lambda[1+3\cos{\left(\theta\right)}-2\lambda]|}{1+\cos{\left(\theta\right)}-\lambda[1+3\cos{\left(\theta\right)}-2\lambda]}.
\end{eqnarray}
Since Eq.~(\ref{eigenvector}) implies that the eigenvectors are determined by the eigenvalues $\{\lambda_j\}_{j=1}^3$, the remaining task is to derive them.
We can show that the three eigenvalues are the solutions of the cubic equation
\begin{eqnarray}
\label{eigenvalue}
\lambda^3-a(\theta,\tilde{\epsilon})\lambda^2+e^{i\pi\tilde{\epsilon}}a^*(\theta,\tilde{\epsilon})\lambda-e^{i\pi\tilde{\epsilon}}=0,
\end{eqnarray}
where $a^*(\theta,\tilde{\epsilon})$ is the complex conjugate of $a(\theta,\tilde{\epsilon})$, and
\begin{eqnarray}
\label{a}
a(\theta,\tilde{\epsilon})\equiv\cfrac{1+3\cos{\left(\theta\right)}}{2}+e^{i\pi\tilde{\epsilon}}\cfrac{1+\cos{\left(\theta\right)}}{2}.
\end{eqnarray}
Cubic equations are solvable due to Cardano's formula, and hence Eq.~(\ref{eigenvalue}) gives us the final state $|\tilde{\psi}(\vec{\epsilon},t)\rangle$ in Eq.~(\ref{finalanalytical}).
The proof of Eqs.~(\ref{eigenvector}) and (\ref{eigenvalue}) is given in Appendix~\ref{A}.

To demonstrate the validity of our approach, we derive the success probability $\sin^2{((2t+1)\theta/2)}$ of the original Grover's algorithm by applying Eqs.~(\ref{eigenvector}) and (\ref{eigenvalue}) with $\tilde{\epsilon}=0$ to Eq.~(\ref{finalanalytical}).
Since $a(\theta,0)=1+2\cos{\left(\theta\right)}$, Eq.~(\ref{eigenvalue}) becomes
\begin{eqnarray}
\lambda^3-\left[1+2\cos{\left(\theta\right)}\right]\lambda(\lambda-1)-1=0,
\end{eqnarray}
and hence $\lambda_1=1$, $\lambda_2=e^{i\theta}$, and $\lambda_3=e^{-i\theta}$.
By substituting these eigenvalues into Eq.~(\ref{eigenvector}), we obtain
\begin{eqnarray}
|\phi_1\rangle=\cfrac{1}{\sqrt{2}}\begin{pmatrix}
0 \\
-1 \\
1 \\
\end{pmatrix},
\end{eqnarray}
\begin{eqnarray}
|\phi_2\rangle=\cfrac{1}{2}\begin{pmatrix}
i\sqrt{2} \\
1 \\
1 \\
\end{pmatrix},
\end{eqnarray}
and
\begin{eqnarray}
|\phi_3\rangle=\cfrac{1}{2}\begin{pmatrix}
-i\sqrt{2} \\
1 \\
1 \\
\end{pmatrix}.
\end{eqnarray}
Therefore, from Eq.~(\ref{finalanalytical}), the final state $|\tilde{\psi}(\vec{\epsilon},t)\rangle$ is
\begin{eqnarray}
\left[D\tilde{O}_f(\tilde{\epsilon})\right]^t|\psi(0)\rangle=\cfrac{i}{\sqrt{2}}\left[-e^{i(t+1/2)\theta}|\phi_2\rangle+e^{-i(t+1/2)\theta}|\phi_3\rangle\right],
\end{eqnarray}
and thus the success probability is
\begin{eqnarray}
\nonumber
&&\left|\langle1'_L|\cfrac{i}{\sqrt{2}}\left[-e^{i(t+1/2)\theta}|\phi_2\rangle+e^{-i(t+1/2)\theta}|\phi_3\rangle\right]\right|^2+\left|\langle1''_L|\cfrac{i}{\sqrt{2}}\left[-e^{i(t+1/2)\theta}|\phi_2\rangle+e^{-i(t+1/2)\theta}|\phi_3\rangle\right]\right|^2\\
&=&\cfrac{\left|1-e^{i(2t+1)\theta}\right|^2}{4}=\sin^2{\left(\cfrac{2t+1}{2}\theta\right)},
\end{eqnarray}
which is the same as that of the original Grover's algorithm.

\begin{figure}[t]
\includegraphics[width=18cm, clip]{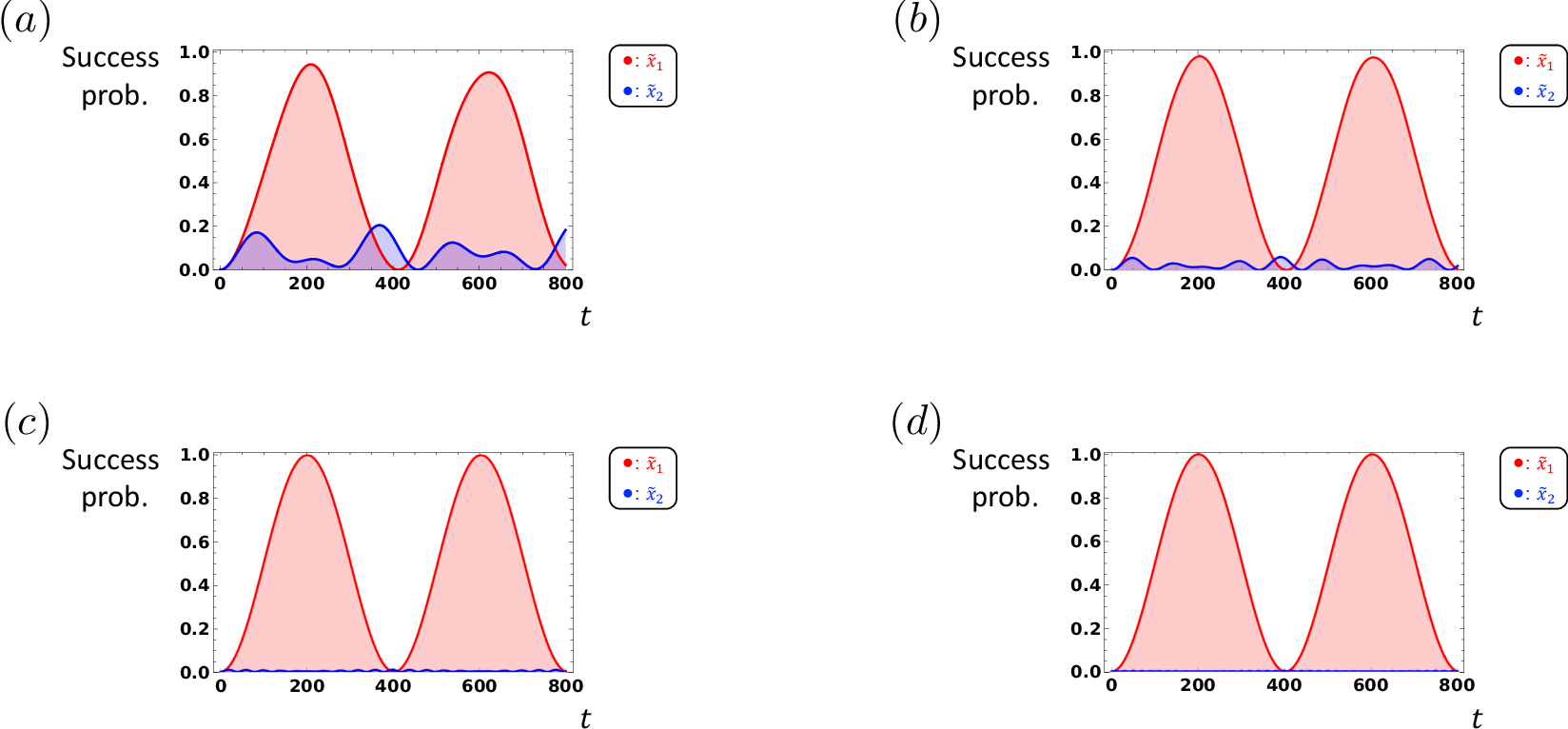}
\caption{Success probabilities of our quantum algorithm. In this figure, we set $n=2^{16}$ and $m=2$. The two marked items in $\mathcal{X}_{1,0}$ and $\mathcal{X}_{1,\tilde{\epsilon}}$ are denoted as $\tilde{x}_1$ and $\tilde{x}_2$, respectively. The red and blue curves represent $P_{\rm suc}^{(\tilde{x}_1)}(\vec{\epsilon},t)$ and $P_{\rm suc}^{(\tilde{x}_2)}(\vec{\epsilon},t)$, respectively. (a) The case of $\tilde{\epsilon}=-0.01$. (b) The case of $\tilde{\epsilon}=-0.02$. (c) The case of $\tilde{\epsilon}=-0.05$. (d) The case of $\tilde{\epsilon}=-0.1$.}
\label{SucProbLfigure}
\end{figure}

The above analysis enables us to perform the numerical simulation with $n=2^{16}$ that is quadratically larger than that in Figs.~\ref{sucprobfigure}, \ref{allsucprobfigure}, and \ref{priorityfigure}.
In Fig.~\ref{SucProbLfigure}\footnote{To confirm the precision of the simulation, we numerically calculate the gap between $\sum_{i=1}^2P_{\rm suc}^{(\tilde{x}_i)}(\vec{\epsilon},0)$ and its correct value $m/n=2^{-15}$ in the range of $-0.1\le\tilde{\epsilon}\le-0.01$. As a result, we obtain $|\sum_{i=1}^2P_{\rm suc}^{(\tilde{x}_i)}(\vec{\epsilon},0)-2^{-15}|<1.3\times 10^{-10}$, which is less than $0.0005\%$ of the correct value $2^{-15}$.}, we plot $P_{\rm suc}^{(\tilde{x}_1)}(\vec{\epsilon},t)$ and $P_{\rm suc}^{(\tilde{x}_2)}(\vec{\epsilon},t)$ for several values of $\tilde{\epsilon}$.
As $\tilde{\epsilon}$ decreases, the first local maxima of $P_{\rm suc}^{(\tilde{x}_1)}(\vec{\epsilon},t)$ and $P_{\rm suc}^{(\tilde{x}_2)}(\vec{\epsilon},t)$ tend to increase and decrease, respectively.
In Appendix~\ref{B}, we analytically approximate $\sum_{x\in\mathcal{X}_{1,0}}P_{\rm suc}^{(x)}(\vec{\epsilon},t)$ and $\sum_{x\in\mathcal{X}_{1,\tilde{\epsilon}}}P_{\rm suc}^{(x)}(\vec{\epsilon},t)$.
As a result, it turns out that these probabilities are $\sin^2{((2t+1)\theta/2)}/2+O(\tilde{\epsilon}^2)$ when $t$ and $\theta$ are fixed [for details, see Eqs.~(\ref{prob1'}) and (\ref{prob1''})].
The original Grover's algorithm is quite robust against the correlated phase errors in the sense that the probabilities do not linearly depend on the noise strength $|\tilde{\epsilon}|$.
Furthermore, in Appendix~\ref{B}, we analytically show that when $|\tilde{\epsilon}|$ is sufficiently small but not $0$, the inequality $\sum_{x\in\mathcal{X}_{1,0}}P_{\rm suc}^{(x)}(\vec{\epsilon},t)>\sum_{x\in\mathcal{X}_{1,\tilde{\epsilon}}}P_{\rm suc}^{(x)}(\vec{\epsilon},t)$ holds by setting $t$ so that (i) $0\le t\theta\ ({\rm mod}\ 2\pi)\le \pi/2$ and (ii) $t>1/\tan{(\theta)}$.
This sufficient condition is consistent with the fact that $P_{\rm suc}^{(\tilde{x}_1)}(\vec{\epsilon},\lfloor\pi/(2\theta)\rfloor)>P_{\rm suc}^{(\tilde{x}_2)}(\vec{\epsilon},\lfloor\pi/(2\theta)\rfloor)$ holds in Fig.~\ref{sucprobfigure}.
Furthermore, it is possible to check whether this condition is satisfied without using values of the priority parameters.

\begin{figure}[t]
\includegraphics[width=7.5cm, clip]{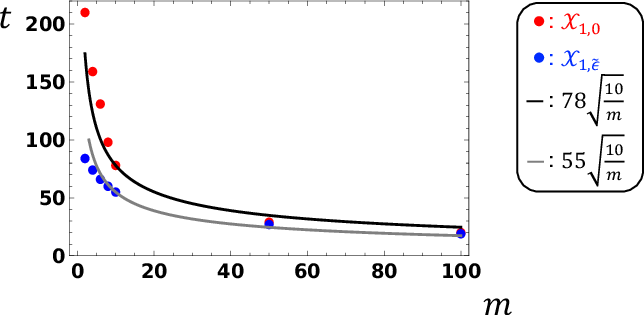}
\caption{The numbers $t$ of queries such that $\sum_{x\in\mathcal{X}_{1,0}}P_{\rm suc}^{(x)}(\vec{\epsilon},t)$ and $\sum_{x\in\mathcal{X}_{1,\tilde{\epsilon}}}P_{\rm suc}^{(x)}(\vec{\epsilon},t)$ take the first local maxima. In this figure, we set $n=2^{16}$ and $\tilde{\epsilon}=-0.01$. The red and blue dots correspond to $\mathcal{X}_{1,0}$ and $\mathcal{X}_{1,\tilde{\epsilon}}$, respectively. The black and gray curves represent $78\sqrt{10/m}$ and $55\sqrt{10/m}$, respectively. Note that these two curves are just guides for the eyes, and thus the coefficients $78\sqrt{10}$ and $55\sqrt{10}$ are meaningless.}
\label{mfigure}
\end{figure}

In Fig.~\ref{mfigure}, we also investigate the dependence of the query complexity of our algorithm on the number $m$ of marked items.
As with the original Grover's algorithm, the query complexity decreases with increasing $m$ and seems to approximately behave as being proportional to $1/\sqrt{m}$ in the range of $2\le m\le 100$.

\subsection{Effect of quantum coherence between marked items}
\label{IIIC}
In this section, we examine the effect of coherence between the marked items in the initial state $|\psi(0)\rangle$.
Coherence is a fundamental property in quantum mechanics and would be necessary to demonstrate quantum advantage.
We consider the same situation as Sec.~\ref{IIIB}.
More specifically, we assume $m=2$, and the two marked items $\tilde{x}_1$ and $\tilde{x}_2$ are in the sets $\mathcal{X}_{1,0}$ and $\mathcal{X}_{1,\tilde{\epsilon}}$, respectively.
We further assume $n=1000$ for simplicity.
In this case, the ideal initial state in Eq.~(\ref{idealini}) is
\begin{eqnarray}
\label{ini2}
|\psi(0)\rangle=\sqrt{\cfrac{499}{500}}|0_L\rangle+\cfrac{1}{\sqrt{1000}}\left(|\tilde{x}_1\rangle+|\tilde{x}_2\rangle\right).
\end{eqnarray}
It is apparent that the state in Eq.~(\ref{ini2}) has the coherence between the first and second marked items.
More formally, when this state is projected onto the space of $\{|\tilde{x}_1\rangle,|\tilde{x}_2\rangle\}$, it becomes $(|\tilde{x}_1\rangle+|\tilde{x}_2\rangle)/\sqrt{2}$.
To quantify its coherence, we use the $l_1$-norm of coherence~\cite{BCP14}.
Let $c_{j,k}$ be the $(j,k)$ element of any density operator $\rho$, i.e., $\rho=\sum_{j,k\in\mathcal{X}}c_{j,k}|j\rangle\langle k|$.
The $l_1$-norm of coherence is defined as
\begin{eqnarray}
C_{l_1}\left(\rho\right)\equiv\sum_{j,k\in\mathcal{X}}|c_{j,k}|-\sum_{j\in\mathcal{X}}|c_{j,j}|.
\end{eqnarray}
By definition, we obtain
\begin{eqnarray}
C_{l_1}\left(\cfrac{|\tilde{x}_1\rangle+|\tilde{x}_2\rangle}{\sqrt{2}}\cfrac{\langle\tilde{x}_1|+\langle\tilde{x}_2|}{\sqrt{2}}\right)=1,
\end{eqnarray}
which is non-zero, and hence $(|\tilde{x}_1\rangle+|\tilde{x}_2\rangle)/\sqrt{2}$ is a coherent state.
We consider how the success probability and query complexity change if the initial state is replaced with the incoherent state
\begin{eqnarray}
\label{rho0}
\rho_0\equiv\cfrac{1}{2}\left[\cfrac{1}{1000}\left(\sqrt{999}|0_L\rangle+|\tilde{x}_1\rangle\right)\left(\sqrt{999}\langle 0_L|+\langle\tilde{x}_1|\right)+\cfrac{1}{1000}\left(\sqrt{999}|0_L\rangle+|\tilde{x}_2\rangle\right)\left(\sqrt{999}\langle 0_L|+\langle\tilde{x}_2|\right)\right].
\end{eqnarray}
Note that $|0_L\rangle$ does not include $|\tilde{x}_1\rangle$ and $|\tilde{x}_2\rangle$ [see Eq.~(\ref{logical0})].
Therefore, $|\tilde{x}_1\rangle$ and $|\tilde{x}_2\rangle$ are not included in the second and first terms, respectively, but they will be produced by Grover's diffusion operator during the algorithm.
When $\rho_0$ is projected onto the space of $\{|\tilde{x}_1\rangle,|\tilde{x}_2\rangle\}$, it becomes $(|\tilde{x}_1\rangle\langle\tilde{x}_1|+|\tilde{x}_2\rangle\langle\tilde{x}_2|)/2$, which is incoherent because
\begin{eqnarray}
C_{l_1}\left(\cfrac{|\tilde{x}_1\rangle\langle\tilde{x}_1|+|\tilde{x}_2\rangle\langle\tilde{x}_2|}{2}\right)=0.
\end{eqnarray}
The reason of why we use $\rho_0$ as an incoherent state is given in Appendix~\ref{C}.

\begin{figure}[t]
\includegraphics[width=18cm, clip]{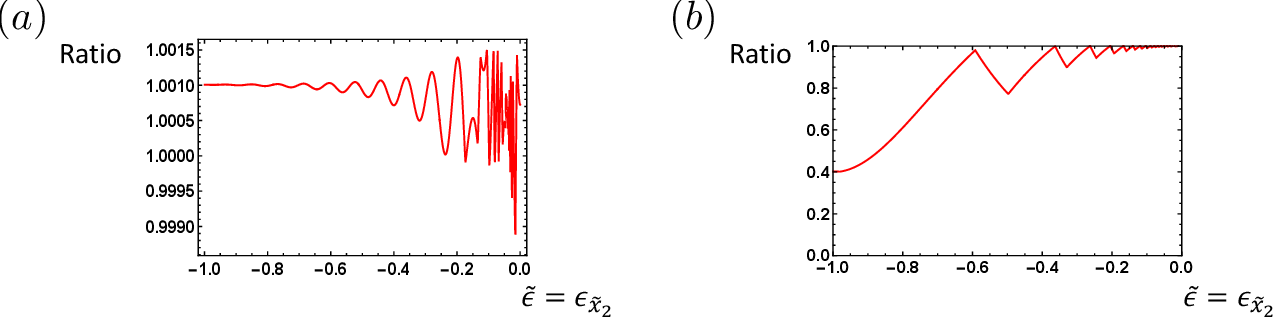}
\caption{Ratios of the first local maximum of the success probability $P_{\rm suc}^{(x)}(\vec{\epsilon},t)$ with the ideal initial state $|\psi(0)\rangle$ to that with the incoherent initial state $\rho_0$. In this figure, we set $n=1000$ and $m=2$ and assume that the first and second marked items $\tilde{x}_1$ and $\tilde{x}_2$ are in the sets $\mathcal{X}_{1,0}$ and $\mathcal{X}_{1,\tilde{\epsilon}}$, respectively. (a) The case of $x=\tilde{x}_1$. That is, the ratio of the probability of finding the first marked item with $|\psi(0)\rangle$ to that with $\rho_0$. (b) The case of $x=\tilde{x}_2$.}
\label{ratiofigure1}
\end{figure}

Recall that $P_{\rm suc}^{(\tilde{x}_i)}(\vec{\epsilon},t)$ is the probability of obtaining the $i$th marked item $\tilde{x}_i$ by measuring $[D\tilde{O}_f(\tilde{\epsilon})]^t|\psi(0)\rangle$ in the basis $\{|x\rangle\}_{x\in\mathcal{X}}$.
Similarly, we define $H_{\rm suc}^{(\tilde{x}_i)}(\vec{\epsilon},t)$ as the probability of obtaining $\tilde{x}_i$ by measuring $[D\tilde{O}_f(\tilde{\epsilon})]^t\rho_0[\tilde{O}_f^\dag(\tilde{\epsilon})D^\dag]^t$ in the same basis.
For any $i$, let $P_{\rm opt}^{(\tilde{x}_i)}(\tilde{\epsilon})$ and $H_{\rm opt}^{(\tilde{x}_i)}(\tilde{\epsilon})$ be the  first local maxima of $P_{\rm suc}^{(\tilde{x}_i)}(\vec{\epsilon},t)$ and $H_{\rm suc}^{(\tilde{x}_i)}(\vec{\epsilon},t)$ over $t$, respectively.
In Fig.~\ref{ratiofigure1}, we plot the ratios $P_{\rm opt}^{(\tilde{x}_1)}(\tilde{\epsilon})/H_{\rm opt}^{(\tilde{x}_1)}(\tilde{\epsilon})$ and $P_{\rm opt}^{(\tilde{x}_2)}(\tilde{\epsilon})/H_{\rm opt}^{(\tilde{x}_2)}(\tilde{\epsilon})$.
Since the ratio in Fig.~\ref{ratiofigure1}(a) is at least $1$ for the most values of $\tilde{\epsilon}$, this figure would imply that the coherence between $\tilde{x}_1$ and $\tilde{x}_2$ tends to increase the success probability of finding the first marked item $\tilde{x}_1$ in our quantum algorithm.
On the other hand, the ratio in Fig.~\ref{ratiofigure1}(b) is at most $1$ for the most values of $\tilde{\epsilon}$, and hence the coherence has the opposite influence on the probability of finding the second marked item $\tilde{x}_2$.

\begin{figure}[t]
\includegraphics[width=18cm, clip]{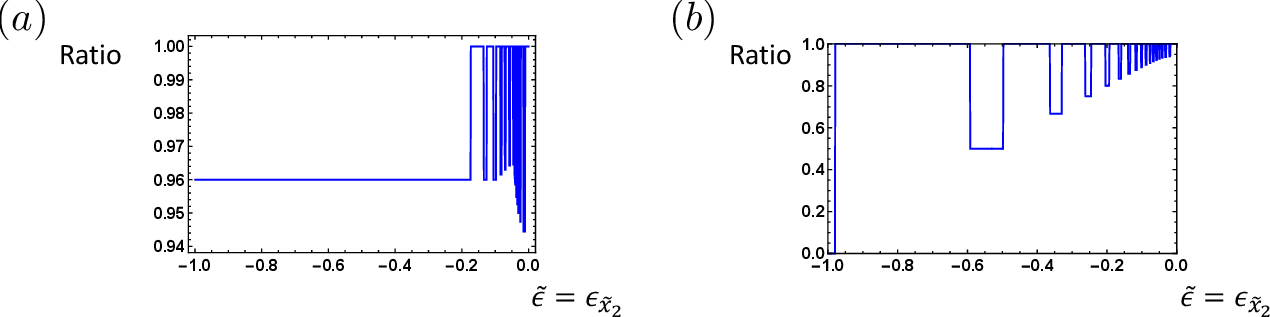}
\caption{Ratios $t_{\rm opt}^{(|\psi(0)\rangle)}(x)/t_{\rm opt}^{(\rho_0)}(x)$. In this figure, we set $n=1000$ and $m=2$ and assume that the first and second marked items $\tilde{x}_1$ and $\tilde{x}_2$ are in the sets $\mathcal{X}_{1,0}$ and $\mathcal{X}_{1,\tilde{\epsilon}}$, respectively. (a) The case of $x=\tilde{x}_1$. That is, the ratio of the smallest argument of the local maximum of the success probability of finding $\tilde{x}_1$ with the initial state $|\psi(0)\rangle$ to that with $\rho_0$.
(b) The case of $x=\tilde{x}_2$.}
\label{ratiofigure3}
\end{figure}

We next consider the query complexity.
For any $i$, let $t_{\rm opt}^{(|\psi(0)\rangle)}(\tilde{x}_i)$ and $t_{\rm opt}^{(\rho_0)}(\tilde{x}_i)$ be the numbers of queries such that $P_{\rm suc}^{(\tilde{x}_i)}(\vec{\epsilon},t_{\rm opt}^{(|\psi(0)\rangle)}(\tilde{x}_i))=P_{\rm opt}^{(\tilde{x}_i)}(\tilde{\epsilon})$ and $H_{\rm suc}^{(\tilde{x}_i)}(\vec{\epsilon},t_{\rm opt}^{(\rho_0)}(\tilde{x}_i))=H_{\rm opt}^{(\tilde{x}_i)}(\tilde{\epsilon})$, respectively.
We numerically calculate the dependence of the ratios $t_{\rm opt}^{(|\psi(0)\rangle)}(\tilde{x}_1)/t_{\rm opt}^{(\rho_0)}(\tilde{x}_1)$ and $t_{\rm opt}^{(|\psi(0)\rangle)}(\tilde{x}_2)/t_{\rm opt}^{(\rho_0)}(\tilde{x}_2)$ on the priority parameter $\tilde{\epsilon}$ in Fig.~\ref{ratiofigure3}.
Unlike Fig.~\ref{ratiofigure1}, the ratio of the query complexity is at most $1$ for all $\tilde{\epsilon}$ owing to the coherence between the two marked items in both cases of $i=1$ and $2$.
This phenomenon would imply that the coherence accelerates our quantum algorithm.

\subsection{Comparison with algorithm in Ref.~\cite{PS08}}
\label{IIID}
In this section, we compare the performance of our quantum algorithm with that of the algorithm in Ref.~\cite{PS08}.
For simplicity, we concretely consider the same situation as Sec.~\ref{IIIB} with $n=8$ and $m=2$.
In this case, the algorithm in Ref.~\cite{PS08} uses the oracle operator
\begin{eqnarray}
\label{oraclePS}
O_f^{({\rm PS})}(\epsilon)\equiv I^{(8)}-2\left(\sqrt{1+\epsilon}|\tilde{x}_1\rangle+\sqrt{-\epsilon}|\tilde{x}_2\rangle\right)
\left(\sqrt{1+\epsilon}\langle\tilde{x}_1|+\sqrt{-\epsilon}\langle\tilde{x}_2|\right)
\end{eqnarray}
with $-1\le\epsilon\le 0$, and thus its final state is $[DO_f^{({\rm PS})}(\epsilon)]^t|\psi(0)\rangle$.
The query complexity $t$ of this algorithm is the closest integer to
\begin{eqnarray}
\cfrac{\cos^{-1}{\left(\sqrt{\cfrac{1+\epsilon}{n}}+\sqrt{-\cfrac{\epsilon}{n}}\right)}}{2\sin^{-1}{\left(\sqrt{\cfrac{1+\epsilon}{n}}+\sqrt{-\cfrac{\epsilon}{n}}\right)}}.
\end{eqnarray}
Therefore, when $n=8$, we can numerically show $t=1$ in the range of $-0.99\le\epsilon\le-0.01$.
As a result, the final state becomes
\begin{eqnarray}
DO_f^{({\rm PS})}(\epsilon)|\psi(0)\rangle=\left(\cfrac{1}{2}-\sqrt{-\epsilon(1+\epsilon)}\right)|\psi(0)\rangle+\left(\sqrt{\cfrac{1+\epsilon}{2}}+\sqrt{-\cfrac{\epsilon}{2}}\right)\left(\sqrt{1+\epsilon}|\tilde{x}_1\rangle+\sqrt{-\epsilon}|\tilde{x}_2\rangle\right).
\end{eqnarray}
By direct calculation, the probabilities $Q_{\rm suc}^{(\tilde{x}_1)}(\epsilon)$ and $Q_{\rm suc}^{(\tilde{x}_2)}(\epsilon)$ of the algorithm in Ref.~\cite{PS08} finding the marked items $\tilde{x}_1$ and $\tilde{x}_2$ are
\begin{eqnarray}
Q_{\rm suc}^{(\tilde{x}_1)}(\epsilon)=\cfrac{\left[1+2\sqrt{-\epsilon(1+\epsilon)}+4(1+\epsilon)\right]^2}{32}
\end{eqnarray}
and
\begin{eqnarray}
Q_{\rm suc}^{(\tilde{x}_2)}(\epsilon)=\cfrac{\left[1+2\sqrt{-\epsilon(1+\epsilon)}-4\epsilon\right]^2}{32},
\end{eqnarray}
respectively.

On the other hand, under the same condition, our oracle operator in Eq.~(\ref{rankedoracle2}) becomes
\begin{eqnarray}
\label{oracleours}
\tilde{O}_f(\tilde{\epsilon})=I^{(8)}-2|\tilde{x}_1\rangle\langle\tilde{x}_1|-(1+e^{i\pi\tilde{\epsilon}})|\tilde{x}_2\rangle\langle\tilde{x}_2|.
\end{eqnarray}
The two parameters $\epsilon$ and $\tilde{\epsilon}$ are not the same, but both are indicators of the priority of $\tilde{x}_1$.
Indeed, the positivity of $(1+2\epsilon)$ or $-\tilde{\epsilon}$ means that $\tilde{x}_1$ is prioritized over $\tilde{x}_2$.
By setting $t=2$, the final state of our algorithm becomes
\begin{eqnarray}
\left[D\tilde{O}_f(\tilde{\epsilon})\right]^2|\psi(0)\rangle=-\cfrac{7+6e^{i\pi\tilde{\epsilon}}+3e^{2i\pi\tilde{\epsilon}}}{16}|\psi(0)\rangle+\cfrac{5-e^{i\pi\tilde{\epsilon}}}{4\sqrt{2}}|\tilde{x}_1\rangle+\cfrac{\left(1+e^{i\pi\tilde{\epsilon}}\right)\left(1+3e^{i\pi\tilde{\epsilon}}\right)}{8\sqrt{2}}|\tilde{x}_2\rangle.
\end{eqnarray}
Here, we define $P_{\rm suc}^{(\tilde{x}_j)}(\tilde{\epsilon})\equiv P_{\rm suc}^{(\tilde{x}_j)}(\vec{\epsilon},2)$ for all $1\le j\le 2$.
The probabilities $P_{\rm suc}^{(\tilde{x}_1)}(\tilde{\epsilon})$ and $P_{\rm suc}^{(\tilde{x}_2)}(\tilde{\epsilon})$ of our algorithm finding the marked items $\tilde{x}_1$ and $\tilde{x}_2$ are
\begin{eqnarray}
P_{\rm suc}^{(\tilde{x}_1)}(\tilde{\epsilon})=\cfrac{373-210\cos{(\tilde{\epsilon}\pi)}-99\cos^2{(\tilde{\epsilon}\pi)}}{512}
\end{eqnarray}
and
\begin{eqnarray}
P_{\rm suc}^{(\tilde{x}_2)}(\tilde{\epsilon})=\cfrac{61+30\cos{(\tilde{\epsilon}\pi)}-27\cos^2{(\tilde{\epsilon}\pi)}}{512},
\end{eqnarray}
respectively.

\begin{figure}[t]
\includegraphics[width=12cm, clip]{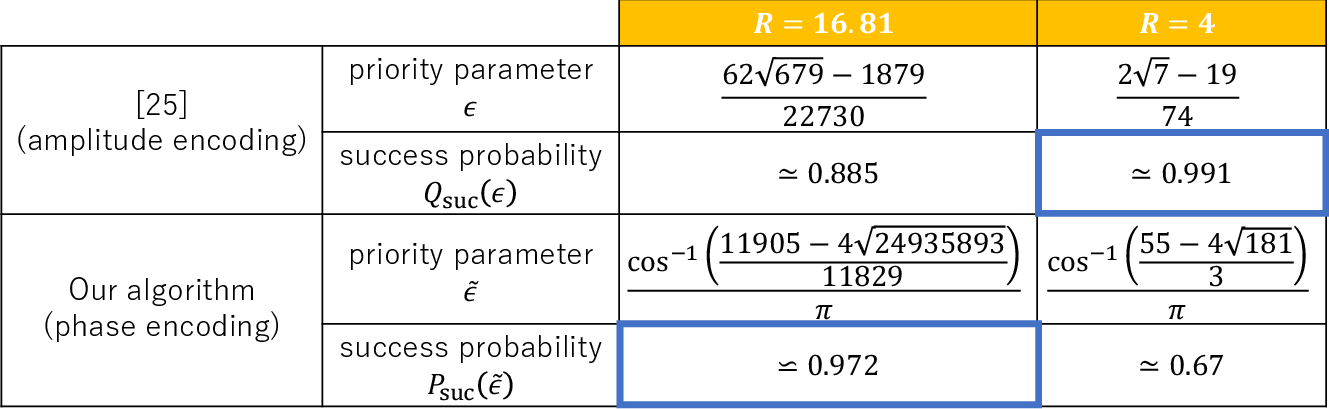}
\caption{Table of parameters used in the numerical simulations in Sec.~\ref{IIID}. The highest success probability in each case of the ratio $R$ is highlighted by the blue rectangle.}
\label{comparisonsummary}
\end{figure}

To compare the two algorithms, we calculate the overall success probabilities $Q_{\rm suc}(\epsilon)\equiv Q_{\rm suc}^{(\tilde{x}_1)}(\epsilon)+Q_{\rm suc}^{(\tilde{x}_2)}(\epsilon)$ and $P_{\rm suc}(\tilde{\epsilon})\equiv P_{\rm suc}^{(\tilde{x}_1)}(\tilde{\epsilon})+P_{\rm suc}^{(\tilde{x}_2)}(\tilde{\epsilon})$ under the following conditions:
\begin{enumerate}
\item[(i)] $-0.99\le\epsilon\le -0.01$
\item[(ii)] $-0.99\le\tilde{\epsilon}\le -0.01$
\item[(iii)] $Q_{\rm suc}^{(\tilde{x}_1)}(\epsilon)/Q_{\rm suc}^{(\tilde{x}_2)}(\epsilon)=P_{\rm suc}^{(\tilde{x}_1)}(\tilde{\epsilon})/P_{\rm suc}^{(\tilde{x}_2)}(\tilde{\epsilon})=R$ for a fixed real value $R$.
\end{enumerate}
When $R=16.81$, $\epsilon=(62\sqrt{679}-1879)/22730$ and $\tilde{\epsilon}=\cos^{-1}{((11905-4\sqrt{24935893})/11829)}/\pi$ hold, and thus $Q_{\rm suc}((62\sqrt{679}-1879)/22730)\simeq0.885$ and $P_{\rm suc}(\cos^{-1}{((11905-4\sqrt{24935893})/11829)}/\pi)\simeq0.972$.
On the other hand, when $R=4$, $\epsilon=(2\sqrt{7}-19)/74$ and $\tilde{\epsilon}=\cos^{-1}{((55-4\sqrt{181})/3)}/\pi$ hold, and thus $Q_{\rm suc}((2\sqrt{7}-19)/74)\simeq0.991$ and $P_{\rm suc}(\cos^{-1}{((55-4\sqrt{181})/3)}/\pi)\simeq0.67$.
They are summarized in Fig.~\ref{comparisonsummary}.
In short, when $R=16.81$ and $R=4$, our algorithm is superior and inferior to the algorithm in Ref.~\cite{PS08}, respectively.
From these observations, we can anticipate that when we would like to achieve a high ratio $R$, our algorithm would be preferred, but when $R$ is sufficiently low, the existing algorithm is preferable.

\begin{figure}[t]
\includegraphics[width=12cm, clip]{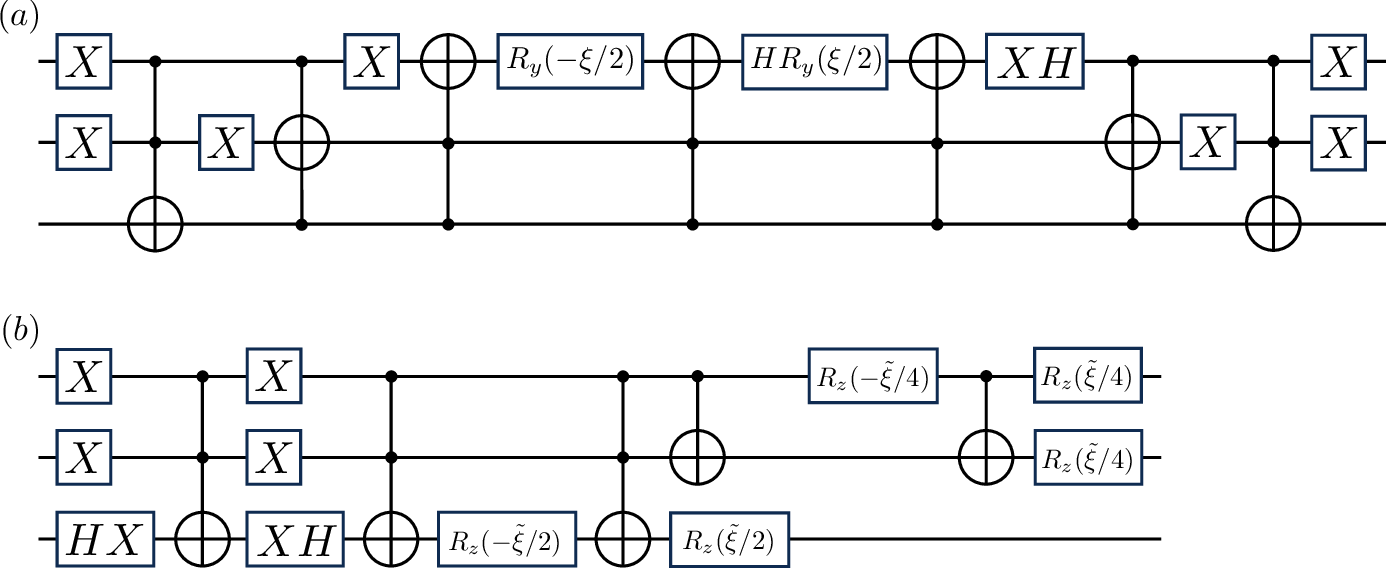}
\caption{Quantum circuits that implement oracle operators. $\xi$ is a real value such that $\cos{(\xi/2)}=-(1+2\epsilon)$ and $\sin{(\xi/2)}=2\sqrt{-\epsilon(1+\epsilon)}$, and $\tilde{\xi}\equiv(1+\tilde{\epsilon})\pi$. $X\equiv|1\rangle\langle 0|+|0\rangle\langle 1|$, $H\equiv\sum_{j,k\in\{0,1\}}(-1)^{jk}|j\rangle\langle k|/\sqrt{2}$, $R_y(\theta)\equiv\cos{(\theta/2)}(|0\rangle\langle 0|+|1\rangle\langle 1|)+\sin{(\theta/2)}(|1\rangle\langle 0|-|0\rangle\langle 1|)$, and $R_z(\theta)\equiv|0\rangle\langle 0|+e^{i\theta}|1\rangle\langle 1|$ for any real value $\theta$. Here, $\{|0\rangle,|1\rangle\}$ are the single-qubit computational basis states. (a) The decomposition of $O_f^{({\rm PS})}(\epsilon)$ in Eq.~(\ref{oraclePS}). (b) The decomposition of $\tilde{O}_f(\tilde{\epsilon})$ in Eq.~(\ref{oracleours}).}
\label{oraclecircuits}
\end{figure}

It would be indispensable for a more detailed comparison to clarify how efficiently these two types of oracle operators can be constructed.
Since we set $n=8$, the marked items can be represented as three-bit strings.
Suppose that $\tilde{x}_1=000$ and $\tilde{x}_2=111$.
We give concrete quantum circuits that implement the oracle operators Eqs.~(\ref{oraclePS}) and (\ref{oracleours}) in Fig.~\ref{oraclecircuits}.
By combining these quantum circuits with the fact that the Toffoli gate can be constructed from single-qubit gates and six CNOT gates~\cite{NC10}, $O_f^{({\rm PS})}(\epsilon)$ and $\tilde{O}_f(\tilde{\epsilon})$ require $42$ and $20$ CNOT gates, respectively.
In this sense, our oracle operator is more efficient than that in Ref.~\cite{PS08}.
On the other hand, although the query complexity $t$ of the algorithm in Ref.~\cite{PS08} is $1$, that of our algorithm is $2$.
Let $|+\rangle\equiv(|0\rangle+|1\rangle)/\sqrt{2}$ with $\{|0\rangle,|1\rangle\}$ being the single-qubit computational basis states.
From 
\begin{eqnarray}
D=-\left[{I^{(2)}}^{\otimes 3}-2\left(|+\rangle\langle +|\right)^{\otimes 3}\right]=-(HX\otimes HX\otimes Z)CCX(XH\otimes XH\otimes Z),
\end{eqnarray}
where $Z=HXH$ and $CCX$ are the Pauli-$Z$ and Toffoli gates, respectively, the implementation of $D$ requires six CNOT gates.
Therefore, the total numbers of CNOT gates (the so-called CNOT count) required to run the algorithm in Ref.~\cite{PS08} and our algorithm are $48$ and $52$, respectively.
Note that the number of CNOT gates in Fig.~\ref{oraclecircuits} may be further reduced by using some optimization method.
It would also be worth mentioning that the more general construction of our oracle operator was explored in Ref.~\cite{SBCJKMMPBS24}.

Lastly, it is worth mentioning that the algorithm in Ref.~\cite{PS08} assumes that the sum of square roots of the priority parameters, which is $\sqrt{1+\epsilon}+\sqrt{-\epsilon}$ in the current situation, is known.
This assumption is different from ours.

\section{Conclusion \& Discussion}
We have generalized Grover's algorithm so that it finds multiple marked items with probabilities according to their priority.
Our quantum algorithm can also be considered as the original Grover's algorithm with correlated phase errors.
We have elaborately analyzed the case where there are two kinds of priority parameters $0$ and $-1\le\tilde{\epsilon}\le 0$.
We have finally compared our quantum algorithm with the existing algorithm~\cite{PS08} and have concluded that which algorithm performs better depends on the priority parameters.

As an outlook, it would be interesting to analyze our quantum algorithm even in the case where there are more than two kinds of priority parameters.
Although it is trivial that our algorithm can be applied to this case, its performance such as the success probability and query complexity is not yet identified.
When there are $k$ kinds of priority 
parameters, our analytical approach in Sec.~\ref{IIIB} requires to solve a $(k+1)$th-degree equation.
Since it is, in general, hard to solve equations with more than fourth degree, the more thorough analysis may necessitate a different approach.
It might be a potential approach to generalize the two-dimensional geometrical interpretation used in Sec.~\ref{II} to a $(k+1)$-dimensional one.
{\red By identifying the query complexity, it becomes possible to examine a quantum advantage of our algorithm.}

In the original Grover's algorithm, if the number $t$ of queries exceeds the optimal value in Eq.~(\ref{queryGrover}), the success probability tragically decreases.
In this sense, the original algorithm needs to know the number $m$ of marked items.
This issue was affirmatively solved by generalizing Grover's algorithm so that it works even if the exact value of $m$ is unknown~\cite{BBHT99,YLC14,CBD23}.
To this end, Ref.~\cite{BBHT99} estimates $m$ before running Grover's algorithm, and Refs.~\cite{YLC14,CBD23} modify the diffusion and oracle operators by introducing phases in them.
It is important to clarify whether the same generalization can be achieved for our quantum algorithm, but it is beyond the scope of this paper.

Recently, several quantum algorithms, which include a quantum algorithm for unstructured search, were understood in a unified way by using the quantum singular value transformation (QSVT)~\cite{MRTC21}.
It would also be interesting to try to understand our quantum algorithm in the framework of the QSVT.
The clarification of whether and how existing theoretical frameworks such as QSVT work for new quantum algorithms will deepen the understanding on quantum computational advantage.

Although we consider only the noise on the oracle operator, other noise models such as the random Gaussian noise on quantum states~\cite{PR99}, the unitary noise on the Hadamard gate~\cite{SMB03}, and the coherent phase noise on (the generalized) Grover's diffusion operator~\cite{LYW23} were also investigated.
To understand the noise robustness of the original Grover's algorithm more deeply, it would be effective to combine our noise model with them.

\section*{DATA AVAILABILITY STATEMENT}
The data cannot be made publicly available because they are not available in a format that is sufficiently accessible or reusable by other researchers.
The data that support the findings of this study are available upon reasonable request from the authors.

\section*{ACKNOWLEDGMENTS}
We thank Suguru Endo for fruitful discussion.
S. Tsuchiya is supported by the Japan Society for the Promotion of Science (JSPS) Grant-in-Aid for Scientific Research (KAKENHI Grant No. 19K03691).
S. Tani is supported by JSPS KAKENHI Grant Numbers JP20H05966 and JP22H00522.
YT is partially supported by JST [Moonshot R\&D -- MILLENNIA Program] Grant Number JPMJMS2061 and MEXT Quantum Leap Flagship Program (MEXT Q-LEAP) Grant Number JPMXS0120319794.

\appendix
\section{\red CONSTRUCTION OF OUR ORACLE OPERATOR FOR GRAPH COLORING PROBLEM (GCP)}
\label{D}
{\red In this appendix, we consider a problem of coloring vertices of a given graph so that two neighboring vertices connected by an edge have different colors.
Since there are, in general, multiple solutions for this problem, suppose that we evaluate the solutions by some reward function $J$.
More formally, a given graph $G$ is represented as $G\equiv(V,E)$ with $V\equiv\{v_i\}_{i=1}^{|V|}$ and $E$ being the sets of vertices and edges, respectively.
We also define the set $C\equiv\{c_i\}_{i=1}^{|C|}$ of available colors and reward matrix $R\equiv(r_{i,j})_{1\le i\le |V|,1\le j\le |C|}$ whose $(i,j)$ element $r_{ij}$ is a non-negative real value and represents the value of the reward given by assigning the $j$th color $c_j$ to the $i$th vertex $v_i$.
Let $A\equiv(a_i)_{1\le i\le |V|}$ be a solution whose $i$th element $a_i$ represens the color that is assinged to $v_i$, i.e., $a_i=j$ when $v_i$ is colored by $c_j$.
The quality of a solution $A$ is evaluated by the reward function
\begin{eqnarray}
\label{RF}
J(A)\equiv\left(\sum_{i=1}^{|V|}r_{i,a_i}\right)\left[\prod_{i,j:(v_i,v_j)\in E}\left(1-\delta_{a_ia_j}\right)\right],
\end{eqnarray}
where the multiplication is taken over all $i$ and $j$ satisfying $(v_i,v_j)\in E$, and $\delta_{a_ia_j}$ is the Kronecker delta.
The problem of finding the best solution $A_{\rm max}\equiv{\rm argmax}_{A}J(A)$ is called the reward-GCP~\cite{SKH22}.}

{\red Specially when the value of $J(A_{\rm max})$ is known, the priority parameter $\epsilon_A$ of our quantum algorithm can be determined by using $J(A)$ as follows:
\begin{eqnarray}
\epsilon_A=-\left(1-\cfrac{J(A)}{J(A_{\rm max})}\right).
\end{eqnarray}
It is easily observed that $\epsilon_A=0$ only when $J(A)=J(A_{\rm max})$, and $\epsilon_A=-1$ if $A$ is not a valid solution, i.e., at least a single pair $(v_i,v_j)\in E$ of neighboring vertices is painted by the same color.
Furthermore, $\epsilon_A>\epsilon_{A'}$ holds when $J(A)>J(A')$, and $-1\le \epsilon_A\le 0$ is satisfied for all $A$ because $r_{i,j}$ is a non-negative real value for all $i$ and $j$.
For simplicity, let us assume that for all $A$, the reward function $J(A)$ can be exactly represented as an $\ell$-bit string, i.e., there exists $j_A^{(1)}j_A^{(2)}\ldots j_A^{(\ell)}\in\{0,1\}^\ell$ such that $J(A)=\sum_{i=1}^\ell j_A^{(i)}2^{i-1}$.
In this case, our oracle operator
\begin{eqnarray}
\tilde{O}_f(\vec{\epsilon})=-\sum_{A}e^{i\pi\epsilon_A}|A\rangle\langle A|=\sum_A e^{i\pi J(A)/J(A_{\rm max})}|A\rangle\langle A|
\end{eqnarray}
can be constructed as follows:
first, we add two ancillary quantum systems $|0^\ell\rangle_{B_1}|0^\ell\rangle_{B_2}$ to the quantum state $|\tilde{\psi}(\vec{\epsilon},t)\rangle_{B_0}=\sum_A\alpha_A|A\rangle_{B_0}$ with $\{\alpha_A\}_A$ being some complex coefficients satisfying $\sum_A|\alpha_A|^2=1$.
Let
\begin{eqnarray}
U_{B_0B_j}\equiv\sum_A|A\rangle\langle A|_{B_0}\otimes\left(\prod_{i=1}^\ell X_i^{j_A^{(i)}}\right)_{B_j}
\end{eqnarray}
be a unitary operator acting on the systems $B_0$ and $B_j$ for any $j\in\{1,2\}$, where $X_i$ is the Pauli-$X$ gate acting on the $i$th qubit in the system $B_j$.
This unitary operator can be performed by coherently calculating $J(A)$ in Eq.~(\ref{RF}).
By using it, we obtain
\begin{eqnarray}
U_{B_0B_2}U_{B_0B_1}|\tilde{\psi}(\vec{\epsilon},t)\rangle_{B_0}|0^\ell\rangle_{B_1}|0^\ell\rangle_{B_2}&=&\sum_A\alpha_A|A\rangle_{B_0}\otimes|j_A^{(1)}j_A^{(2)}\ldots j_A^{(\ell)}\rangle_{B_1}\otimes|j_A^{(1)}j_A^{(2)}\ldots j_A^{(\ell)}\rangle_{B_2}\\
&=&\sum_A\alpha_A|A\rangle_{B_0}\otimes|B(J(A))\rangle_{B_1}|B(J(A))\rangle_{B_2},
\end{eqnarray}
where $B(J(A))$ is the binary representation of $J(A)$.
Then, for all $1\le k\le\ell$, we apply $\Lambda(R_z(2^{k-1}\pi/J(A_{\rm max})))\equiv|0\rangle\langle 0|\otimes I^{(2)}+|1\rangle\langle 1|\otimes R_z(2^{k-1}\pi/J(A_{\rm max}))$ on the $k$th qubits in the systems $B_1$ and $B_2$, where $R_z(\theta)\equiv|0\rangle\langle 0|+e^{i\theta}|1\rangle\langle 1|$ for any real value $\theta$.
As a result, the quantum state becomes
\begin{eqnarray}
\sum_A\alpha_Ae^{i\pi J(A)/J(A_{\rm max})}|A\rangle_{B_0}\otimes|B(J(A))\rangle_{B_1}|B(J(A))\rangle_{B_2}
\end{eqnarray}
because $\Lambda(R_z(2^{k-1}\pi/J(A_{\rm max})))|j_A^{(k)}\rangle|j_A^{(k)}\rangle=e^{i\pi j_A^{(k)}2^{k-1}/J(A_{\rm max})}|j_A^{(k)}\rangle|j_A^{(k)}\rangle$.
Finally, by applying $U_{B_0B_1}^\dag U_{B_0B_2}^\dag$ and discarding the systems $B_1$ and $B_2$, we can obtain
\begin{eqnarray}
\sum_A\alpha_Ae^{i\pi J(A)/J(A_{\rm max})}|A\rangle=\tilde{O}_f(\vec{\epsilon})|\tilde{\psi}(\vec{\epsilon},t)\rangle.
\end{eqnarray}}

\section{DERIVATION OF EQS.~(\ref{eigenvector}) AND (\ref{eigenvalue})}
\label{A}
Let $|\phi\rangle=\alpha|0_L\rangle+\beta|1'_L\rangle+\gamma|1''_L\rangle$ be a normalized eigenvector of $D\tilde{O}_f(\tilde{\epsilon})$ associated with the eigenvalue $\lambda$.
Since we can assume that $\gamma$ is real without loss of generality, $|\alpha|^2+|\beta|^2+\gamma^2=1$.
From $D\tilde{O}_f(\tilde{\epsilon})|\phi\rangle=\lambda|\phi\rangle$, we obtain the following system of equations:
\begin{align}
\label{A1}
\left\{
\begin{array}{lll}
\left[\cos{\left(\theta\right)}-\lambda\right]\alpha-\cfrac{\sin{\left(\theta\right)}}{\sqrt{2}}\beta-e^{i\pi\tilde{\epsilon}}\cfrac{\sin{\left(\theta\right)}}{\sqrt{2}}\gamma&=&0 \\
\cfrac{\sin{\left(\theta\right)}}{\sqrt{2}}\alpha+\left[\cfrac{1+\cos{\left(\theta\right)}}{2}-\lambda\right]\beta-e^{i\pi\tilde{\epsilon}}\cfrac{1-\cos{\left(\theta\right)}}{2}\gamma&=&0 \\
\cfrac{\sin{\left(\theta\right)}}{\sqrt{2}}\alpha-\cfrac{1-\cos{\left(\theta\right)}}{2}\beta+\left[e^{i\pi\tilde{\epsilon}}\cfrac{1+\cos{\left(\theta\right)}}{2}-\lambda\right]\gamma&=&0.
\end{array}
\right.
\end{align}
Since $D\tilde{O}_f(\tilde{\epsilon})$ is a unitary operator, $|\lambda|=1$.
On the other hand, from $1\le m\le n/2$, we have $0\le\cos{\left(\theta\right)}<1$ because $\cos{\left(\theta\right)}=1-2\sin^2{\left(\theta/2\right)}=1-2m/n$.
Therefore, $\lambda\neq\cos{\left(\theta\right)}$, and hence the first equality in Eq.~(\ref{A1}) implies
\begin{eqnarray}
\label{alpha}
\alpha=\cfrac{\sin{\left(\theta\right)}\left(\beta+e^{i\pi\tilde{\epsilon}}\gamma\right)}{\sqrt{2}\left[\cos{\left(\theta\right)}-\lambda\right]}.
\end{eqnarray}
By substituting Eq.~(\ref{alpha}) into the second equality in Eq.~(\ref{A1}),
\begin{eqnarray}
&&\cfrac{\sin{\left(\theta\right)}}{\sqrt{2}}\cfrac{\sin{\left(\theta\right)}\left(\beta+e^{i\pi\tilde{\epsilon}}\gamma\right)}{\sqrt{2}\left[\cos{\left(\theta\right)}-\lambda\right]}+\left[\cfrac{1+\cos{\left(\theta\right)}}{2}-\lambda\right]\beta-e^{i\pi\tilde{\epsilon}}\cfrac{1-\cos{\left(\theta\right)}}{2}\gamma=0\\
&\Rightarrow&\cfrac{\sin^2{\left(\theta\right)}+\left[1+\cos{\left(\theta\right)}-2\lambda\right]\left[\cos{\left(\theta\right)}-\lambda\right]}{2\left[\cos{\left(\theta\right)}-\lambda\right]}\beta+e^{i\pi\tilde{\epsilon}}\cfrac{\sin^2{\left(\theta\right)}-\left[1-\cos{\left(\theta\right)}\right]\left[\cos{\left(\theta\right)}-\lambda\right]}{2\left[\cos{\left(\theta\right)}-\lambda\right]}\gamma=0\\
\label{A5}
&\Rightarrow&\cfrac{\left(1-\lambda\right)\left[1+\cos{\left(\theta\right)}\right]+2\lambda\left[\lambda-\cos{\left(\theta\right)}\right]}{2\left[\cos{\left(\theta\right)}-\lambda\right]}\beta+e^{i\pi\tilde{\epsilon}}\cfrac{\left(1+\lambda\right)\left[1-\cos{\left(\theta\right)}\right]}{2\left[\cos{\left(\theta\right)}-\lambda\right]}\gamma=0.
\end{eqnarray}

To prove that the coefficient of $\beta$ in Eq.~(\ref{A5}) is not $0$, we show $\left(1-\lambda\right)\left[1+\cos{\left(\theta\right)}\right]+2\lambda\left[\lambda-\cos{\left(\theta\right)}\right]\neq0$ for any $\lambda$ and $\theta$ under the condition $0\le\cos{\left(\theta\right)}<1$.
If $\left(1-\lambda\right)\left[1+\cos{\left(\theta\right)}\right]+2\lambda\left[\lambda-\cos{\left(\theta\right)}\right]=0$, then
\begin{eqnarray}
&&\lambda=\cfrac{1+3\cos{\left(\theta\right)}\pm\sqrt{\left[1+3\cos{\left(\theta\right)}\right]^2-8\left[1+\cos{\left(\theta\right)}\right]}}{4}=\cfrac{1+3\cos{\left(\theta\right)}\pm i\sqrt{\left[1-\cos{\left(\theta\right)}\right]\left[7+9\cos{\left(\theta\right)}\right]}}{4}\\
&\Rightarrow&|\lambda|^2=\cfrac{\left[1+3\cos{\left(\theta\right)}\right]^2+\left[1-\cos{\left(\theta\right)}\right]\left[7+9\cos{\left(\theta\right)}\right]}{16}=\cfrac{1+\cos{\left(\theta\right)}}{2}.
\end{eqnarray}
Therefore, $|\lambda|\neq1$, which contradicts to the fact that $D\tilde{O}_f(\tilde{\epsilon})$ is a unitary operator.
In conclusion, $\left(1-\lambda\right)\left[1+\cos{\left(\theta\right)}\right]+2\lambda\left[\lambda-\cos{\left(\theta\right)}\right]\neq0$.
Now, we can immediately obtain
\begin{eqnarray}
\label{beta}
\beta=-e^{i\pi\tilde{\epsilon}}\cfrac{(1+\lambda)\left[1-\cos{\left(\theta\right)}\right]}{(1-\lambda)\left[1+\cos{\left(\theta\right)}\right]+2\lambda\left[\lambda-\cos{\left(\theta\right)}\right]}\gamma
\end{eqnarray}
from Eq.~(\ref{A5}).
By substituting Eq.~(\ref{beta}) into Eq.~(\ref{alpha}),
\begin{eqnarray}
\alpha&=&\cfrac{\sin{\left(\theta\right)}}{\sqrt{2}\left[\cos{\left(\theta\right)}-\lambda\right]}\left\{-e^{i\pi\tilde{\epsilon}}\cfrac{(1+\lambda)\left[1-\cos{\left(\theta\right)}\right]}{(1-\lambda)\left[1+\cos{\left(\theta\right)}\right]+2\lambda\left[\lambda-\cos{\left(\theta\right)}\right]}+e^{i\pi\tilde{\epsilon}}\right\}\gamma\\
\label{alpha2}
&=&e^{i\pi\tilde{\epsilon}}\cfrac{\sqrt{2}(1-\lambda)\sin{\left(\theta\right)}}{(1-\lambda)\left[1+\cos{\left(\theta\right)}\right]+2\lambda\left[\lambda-\cos{\left(\theta\right)}\right]}\gamma.
\end{eqnarray}
By using Eqs.~(\ref{beta}) and (\ref{alpha2}),
\begin{eqnarray}
&&|\alpha|^2+|\beta|^2+\gamma^2=1\\
&\Rightarrow&\cfrac{2|1-\lambda|^2\sin^2{\left(\theta\right)}}{|(1-\lambda)\left[1+\cos{\left(\theta\right)}\right]+2\lambda\left[\lambda-\cos{\left(\theta\right)}\right]|^2}\gamma^2+\cfrac{|1+\lambda|^2\left[1-\cos{\left(\theta\right)}\right]^2}{|(1-\lambda)\left[1+\cos{\left(\theta\right)}\right]+2\lambda\left[\lambda-\cos{\left(\theta\right)}\right]|^2}\gamma^2+\gamma^2=1\\
&\Rightarrow&\gamma=\cfrac{|(1-\lambda)\left[1+\cos{\left(\theta\right)}\right]+2\lambda\left[\lambda-\cos{\left(\theta\right)}\right]|}{\sqrt{2|1-\lambda|^2\sin^2{\left(\theta\right)}+|1+\lambda|^2\left[1-\cos{\left(\theta\right)}\right]^2+|(1-\lambda)\left[1+\cos{\left(\theta\right)}\right]+2\lambda\left[\lambda-\cos{\left(\theta\right)}\right]|^2}}\\
\label{gamma}
&\Rightarrow&\gamma=\cfrac{|1+\cos{\left(\theta\right)}-\lambda\left[1+3\cos{\left(\theta\right)}-2\lambda\right]|}{\sqrt{2|1-\lambda|^2\sin^2{\left(\theta\right)}+|1+\lambda|^2\left[1-\cos{\left(\theta\right)}\right]^2+|1+\cos{\left(\theta\right)}-\lambda\left[1+3\cos{\left(\theta\right)}-2\lambda\right]|^2}}.
\end{eqnarray}
Eq.~(\ref{gamma}) together with Eqs.~(\ref{beta}) and (\ref{alpha2}) implies our objective Eq.~(\ref{eigenvector}).

We next derive Eq.~(\ref{eigenvalue}).
It is obtained from the characteristic equation
\begin{eqnarray}
\label{eigenformula}
{\rm det}\left(D\tilde{O}_f(\tilde{\epsilon})-\lambda I^{(3)}\right)=0.
\end{eqnarray}
The left-hand side of Eq.~(\ref{eigenformula}) is
\begin{eqnarray}
\nonumber
&&\left[\cos{\left(\theta\right)}-\lambda\right]\left[\cfrac{1+\cos{\left(\theta\right)}}{2}-\lambda\right]\left[e^{i\pi\tilde{\epsilon}}\cfrac{1+\cos{\left(\theta\right)}}{2}-\lambda\right]+e^{i\pi\tilde{\epsilon}}\sin^2{\left(\theta\right)}\cfrac{1-\cos{\left(\theta\right)}}{2}+e^{i\pi\tilde{\epsilon}}\cfrac{\sin^2{\left(\theta\right)}}{2}\left[\cfrac{1+\cos{\left(\theta\right)}}{2}-\lambda\right]\\
&&+\cfrac{\sin^2{\left(\theta\right)}}{2}\left[e^{i\pi\tilde{\epsilon}}\cfrac{1+\cos{\left(\theta\right)}}{2}-\lambda\right]-e^{i\pi\tilde{\epsilon}}\left[\cfrac{1-\cos{\left(\theta\right)}}{2}\right]^2\left[\cos{\left(\theta\right)}-\lambda\right]\\
\nonumber
&=&\left[\lambda^2-\cfrac{1+3\cos{\left(\theta\right)}}{2}\lambda+\cos{\left(\theta\right)}\cfrac{1+\cos{\left(\theta\right)}}{2}\right]\left[e^{i\pi\tilde{\epsilon}}\cfrac{1+\cos{\left(\theta\right)}}{2}-\lambda\right]+e^{i\pi\tilde{\epsilon}}\sin^2{\left(\theta\right)}\cfrac{1-\cos{\left(\theta\right)}}{2}+e^{i\pi\tilde{\epsilon}}\sin^2{\left(\theta\right)}\cfrac{1+\cos{\left(\theta\right)}}{2}\\
&&-e^{i\pi\tilde{\epsilon}}\cfrac{\sin^2{\left(\theta\right)}}{2}\lambda-\cfrac{\sin^2{\left(\theta\right)}}{2}\lambda-e^{i\pi\tilde{\epsilon}}\left[\cfrac{1-\cos{\left(\theta\right)}}{2}\right]^2\cos{\left(\theta\right)}+e^{i\pi\tilde{\epsilon}}\left[\cfrac{1-\cos{\left(\theta\right)}}{2}\right]^2\lambda\\
\nonumber
&=&-\lambda^3+\left[\cfrac{1+3\cos{\left(\theta\right)}}{2}+e^{i\pi\tilde{\epsilon}}\cfrac{1+\cos{\left(\theta\right)}}{2}\right]\lambda^2-\left[\cos{\left(\theta\right)}\cfrac{1+\cos{\left(\theta\right)}}{2}+e^{i\pi\tilde{\epsilon}}\cfrac{1+3\cos{\left(\theta\right)}}{2}\cfrac{1+\cos{\left(\theta\right)}}{2}\right]\lambda\\
\nonumber
&&+e^{i\pi\tilde{\epsilon}}\cos{\left(\theta\right)}\left[\cfrac{1+\cos{\left(\theta\right)}}{2}\right]^2+e^{i\pi\tilde{\epsilon}}\sin^2{\left(\theta\right)}-\left(1+e^{i\pi\tilde{\epsilon}}\right)\cfrac{\sin^2{\left(\theta\right)}}{2}\lambda-e^{i\pi\tilde{\epsilon}}\left[\cfrac{1-\cos{\left(\theta\right)}}{2}\right]^2\cos{\left(\theta\right)}+e^{i\pi\tilde{\epsilon}}\left[\cfrac{1-\cos{\left(\theta\right)}}{2}\right]^2\lambda\\
&&\\
\nonumber
&=&-\lambda^3+\left[\cfrac{1+3\cos{\left(\theta\right)}}{2}+e^{i\pi\tilde{\epsilon}}\cfrac{1+\cos{\left(\theta\right)}}{2}\right]\lambda^2\\
\nonumber
&&-\left\{\cos{\left(\theta\right)}\cfrac{1+\cos{\left(\theta\right)}}{2}+e^{i\pi\tilde{\epsilon}}\left[\cfrac{3\cos^2{\left(\theta\right)}+4\cos{\left(\theta\right)}+1}{4}-\cfrac{\cos^2{\left(\theta\right)}-2\cos{\left(\theta\right)}+1}{4}\right]\right\}\lambda\\
&&+e^{i\pi\tilde{\epsilon}}\cos{\left(\theta\right)}\left[\cfrac{\cos^2{\left(\theta\right)}+2\cos{\left(\theta\right)}+1}{4}-\cfrac{\cos^2{\left(\theta\right)}-2\cos{\left(\theta\right)}+1}{4}\right]+e^{i\pi\tilde{\epsilon}}\sin^2{\left(\theta\right)}-\left(1+e^{i\pi\tilde{\epsilon}}\right)\cfrac{\sin^2{\left(\theta\right)}}{2}\lambda\\
&=&-\lambda^3+\left[\cfrac{1+3\cos{\left(\theta\right)}}{2}+e^{i\pi\tilde{\epsilon}}\cfrac{1+\cos{\left(\theta\right)}}{2}\right]\lambda^2-\left[\cfrac{\cos^2{\left(\theta\right)}+\cos{\left(\theta\right)}}{2}+\cfrac{\sin^2{\left(\theta\right)}}{2}+e^{i\pi\tilde{\epsilon}}\cfrac{1+3\cos{\left(\theta\right)}}{2}\right]\lambda+e^{i\pi\tilde{\epsilon}}\\
&=&-\lambda^3+\left[\cfrac{1+3\cos{\left(\theta\right)}}{2}+e^{i\pi\tilde{\epsilon}}\cfrac{1+\cos{\left(\theta\right)}}{2}\right]\lambda^2-\left[\cfrac{1+\cos{\left(\theta\right)}}{2}+e^{i\pi\tilde{\epsilon}}\cfrac{1+3\cos{\left(\theta\right)}}{2}\right]\lambda+e^{i\pi\tilde{\epsilon}}\\
\label{eigenformulaf}
&=&-\lambda^3+a(\theta,\tilde{\epsilon})\lambda^2-e^{i\pi\tilde{\epsilon}}a^\ast(\theta,\tilde{\epsilon})\lambda+e^{i\pi\tilde{\epsilon}},
\end{eqnarray}
where $a(\theta,\tilde{\epsilon})$ is defined in Eq.~(\ref{a}).
By substituting Eq.~(\ref{eigenformulaf}) into Eq.~(\ref{eigenformula}), we immediately obtain Eq.~(\ref{eigenvalue}).

\section{APPROXIMATION OF SUCCESS PROBABILITIES OF OUR ALGORITHM}
\label{B}
Under the same situation as Sec.~\ref{IIIB} with $1\le m\le n/2$, we derive the approximation of the success probabilities $\sum_{x\in\mathcal{X}_{1,0}}P_{\rm suc}^{(x)}(\vec{\epsilon},t)$ and $\sum_{x\in\mathcal{X}_{1,\tilde{\epsilon}}}P_{\rm suc}^{(x)}(\vec{\epsilon},t)$ of our quantum algorithm.
To this end, we transform the orthonormal basis from $\{|0_L\rangle,|1'_L\rangle,|1''_L\rangle\}$ to $\{|0_L\rangle,|+_L\rangle,|-_L\rangle\}$ with $|\pm_L\rangle\equiv(|1'_L\rangle\pm|1''_L\rangle)/\sqrt{2}$.
From Eq.~(\ref{DOmf}), the matrix form of $D\tilde{O}_f(\tilde{\epsilon})$ in this transformed basis is
\begin{eqnarray}
\begin{pmatrix}
\cos{\left(\theta\right)} & -\cfrac{1+e^{i\pi\tilde{\epsilon}}}{2}\sin{\left(\theta\right)} & -\cfrac{1-e^{i\pi\tilde{\epsilon}}}{2}\sin{\left(\theta\right)} \\
\sin{\left(\theta\right)} & \cfrac{1+e^{i\pi\tilde{\epsilon}}}{2}\cos{\left(\theta\right)} & \cfrac{1-e^{i\pi\tilde{\epsilon}}}{2}\cos{\left(\theta\right)} \\
0 & \cfrac{1-e^{i\pi\tilde{\epsilon}}}{2} & \cfrac{1+e^{i\pi\tilde{\epsilon}}}{2} \\
\end{pmatrix}
=
\begin{pmatrix}
\cos{\left(\theta\right)} & -\sin{\left(\theta\right)} & 0 \\
\sin{\left(\theta\right)} & \cos{\left(\theta\right)} & 0 \\
0 & 0 & 1 \\
\end{pmatrix}
+\cfrac{1-e^{i\pi\tilde{\epsilon}}}{2}
\begin{pmatrix}
0 & \sin{\left(\theta\right)} & -\sin{\left(\theta\right)} \\
0 & -\cos{\left(\theta\right)} & \cos{\left(\theta\right)} \\
0 & 1 & -1 \\
\end{pmatrix}.
\end{eqnarray}
Let
\begin{eqnarray}
G\equiv\begin{pmatrix}
\cos{\left(\theta\right)} & -\sin{\left(\theta\right)} & 0 \\
\sin{\left(\theta\right)} & \cos{\left(\theta\right)} & 0 \\
0 & 0 & 1 \\
\end{pmatrix}
\end{eqnarray}
and
\begin{eqnarray}
E\equiv\begin{pmatrix}
0 & \sin{\left(\theta\right)} & -\sin{\left(\theta\right)} \\
0 & -\cos{\left(\theta\right)} & \cos{\left(\theta\right)} \\
0 & 1 & -1 \\
\end{pmatrix}.
\end{eqnarray}
When $|\tilde{\epsilon}|$ is sufficiently small, $(1-e^{i\pi\tilde{\epsilon}})/2\simeq-i\pi\tilde{\epsilon}/2+\pi^2\tilde{\epsilon}^2/4$, and hence
\begin{eqnarray}
\label{appDO}
\left[D\tilde{O}_f(\tilde{\epsilon})\right]^t\simeq G^t+\left(-i\cfrac{\pi}{2}\tilde{\epsilon}+\cfrac{\pi^2}{4}\tilde{\epsilon}^2\right)\sum_{j=0}^{t-1}G^jEG^{t-j-1}-\cfrac{\pi^2}{4}\tilde{\epsilon}^2\sum_{j=0}^{t-2}\sum_{k=0}^{t-j-2}G^jEG^kEG^{t-j-k-2}.
\end{eqnarray}
Note that in Eq.~(\ref{appDO}), when $t<2$, the third term is zero, and when $t=0$, the second term is also zero.

To achieve our purpose, it is sufficient to calculate $|\langle 1'_L|[D\tilde{O}_f(\tilde{\epsilon})]^t|\psi(0)\rangle|^2$ and $|\langle 1''_L|[D\tilde{O}_f(\tilde{\epsilon})]^t|\psi(0)\rangle|^2$.
To this end, we first derive the matrix form of $G^lE$ and calculate $G^l|\psi(0)\rangle$ for any $0\le l\le t$.
Since $G$ is a rotation matrix, we obtain
\begin{eqnarray}
\label{GE}
G^lE=\begin{pmatrix}
\cos{\left(l\theta\right)} & -\sin{\left(l\theta\right)} & 0 \\
\sin{\left(l\theta\right)} & \cos{\left(l\theta\right)} & 0 \\
0 & 0 & 1 \\
\end{pmatrix}
\begin{pmatrix}
0 & \sin{\left(\theta\right)} & -\sin{\left(\theta\right)} \\
0 & -\cos{\left(\theta\right)} & \cos{\left(\theta\right)} \\
0 & 1 & -1 \\
\end{pmatrix}
=
\begin{pmatrix}
0 & \sin{\left(\left(l+1\right)\theta\right)} & -\sin{\left(\left(l+1\right)\theta\right)} \\
0 & -\cos{\left(\left(l+1\right)\theta\right)} & \cos{\left(\left(l+1\right)\theta\right)} \\
0 & 1 & -1 \\
\end{pmatrix}
\end{eqnarray}
and
\begin{eqnarray}
\label{Gpsi}
G^l|\psi(0)\rangle=\begin{pmatrix}
\cos{\left(l\theta\right)} & -\sin{\left(l\theta\right)} & 0 \\
\sin{\left(l\theta\right)} & \cos{\left(l\theta\right)} & 0 \\
0 & 0 & 1 \\
\end{pmatrix}
\begin{pmatrix}
\cos{\left(\cfrac{\theta}{2}\right)} \\
\sin{\left(\cfrac{\theta}{2}\right)} \\
0 \\
\end{pmatrix}
=
\begin{pmatrix}
\cos{\left(\left(l+\cfrac{1}{2}\right)\theta\right)} \\
\sin{\left(\left(l+\cfrac{1}{2}\right)\theta\right)} \\
0 \\
\end{pmatrix}.
\end{eqnarray}
From Eqs.~(\ref{appDO}), (\ref{GE}), and (\ref{Gpsi}),
\begin{eqnarray}
\nonumber
&&\langle 1'_L|[D\tilde{O}_f(\tilde{\epsilon})]^t|\psi(0)\rangle\\
\nonumber
&\simeq&\cfrac{1}{\sqrt{2}}\begin{pmatrix}
0 & 1 & 1 \\
\end{pmatrix}
\left[\begin{pmatrix}
\cos{\left(\left(t+\cfrac{1}{2}\right)\theta\right)} \\
\sin{\left(\left(t+\cfrac{1}{2}\right)\theta\right)} \\
0 \\
\end{pmatrix}\right.\\
\nonumber
&&+\left(-i\cfrac{\pi}{2}\tilde{\epsilon}+\cfrac{\pi^2}{4}\tilde{\epsilon}^2\right)\sum_{j=0}^{t-1}\begin{pmatrix}
0 & \sin{\left(\left(j+1\right)\theta\right)} & -\sin{\left(\left(j+1\right)\theta\right)} \\
0 & -\cos{\left(\left(j+1\right)\theta\right)} & \cos{\left(\left(j+1\right)\theta\right)} \\
0 & 1 & -1 \\
\end{pmatrix}
\begin{pmatrix}
\cos{\left(\left(t-j-\cfrac{1}{2}\right)\theta\right)} \\
\sin{\left(\left(t-j-\cfrac{1}{2}\right)\theta\right)} \\
0 \\
\end{pmatrix}\\
&&+\cfrac{\pi^2}{4}\tilde{\epsilon}^2\sum_{j=0}^{t-2}\sum_{k=0}^{t-j-2}\left[1+\cos{\left((k+1)\theta\right)}\right]\left.\begin{pmatrix}
0 & \sin{\left(\left(j+1\right)\theta\right)} & -\sin{\left(\left(j+1\right)\theta\right)} \\
0 & -\cos{\left(\left(j+1\right)\theta\right)} & \cos{\left(\left(j+1\right)\theta\right)} \\
0 & 1 & -1 \\
\end{pmatrix}
\begin{pmatrix}
\cos{\left(\left(t-j-k-\cfrac{3}{2}\right)\theta\right)} \\
\sin{\left(\left(t-j-k-\cfrac{3}{2}\right)\theta\right)} \\
0 \\
\end{pmatrix}\right]\ \ \ \ \ \\
\nonumber
&=&\cfrac{\sin{\left(\left(t+\cfrac{1}{2}\right)\theta\right)}}{\sqrt{2}}+\left(-i\cfrac{\pi}{2}\tilde{\epsilon}+\cfrac{\pi^2}{4}\tilde{\epsilon}^2\right)\sum_{j=0}^{t-1}\cfrac{1-\cos{\left(\left(j+1\right)\theta\right)}}{\sqrt{2}}\sin{\left(\left(t-j-\cfrac{1}{2}\right)\theta\right)}\\
\label{amp1}
&&+\cfrac{\pi^2}{4}\tilde{\epsilon}^2\sum_{j=0}^{t-2}\sum_{k=0}^{t-j-2}\left[1+\cos{\left(\left(k+1\right)\theta\right)}\right]\cfrac{1-\cos{\left(\left(j+1\right)\theta\right)}}{\sqrt{2}}\sin{\left(\left(t-j-k-\cfrac{3}{2}\right)\theta\right)}.
\end{eqnarray}
For simplicity, we define
\begin{eqnarray}
s_1&\equiv&\sum_{j=0}^{t-1}\left[1-\cos{\left(\left(j+1\right)\theta\right)}\right]\sin{\left(\left(t-j-\cfrac{1}{2}\right)\theta\right)}\\
&=&\cfrac{\sin^2{\left(\cfrac{t\theta}{2}\right)}}{\sin{\left(\cfrac{\theta}{2}\right)}}-\cfrac{1}{2}\sum_{j=0}^{t-1}\left[\sin{\left(\left(t+\cfrac{1}{2}\right)\theta\right)}+\sin{\left(\left(t-2j-\cfrac{3}{2}\right)\theta\right)}\right]\\
&=&\cfrac{\sin^2{\left(\cfrac{t\theta}{2}\right)}}{\sin{\left(\cfrac{\theta}{2}\right)}}-\cfrac{t}{2}\sin{\left(\left(t+\cfrac{1}{2}\right)\theta\right)}+\cfrac{\sin{\left(t\theta\right)}\sin{\left(\cfrac{\theta}{2}\right)}}{2\sin{\left(\theta\right)}}
\end{eqnarray}
and
\begin{eqnarray}
s_2\equiv\sum_{j=0}^{t-2}\sum_{k=0}^{t-j-2}\left[1+\cos{\left(\left(k+1\right)\theta\right)}\right]\left[1-\cos{\left(\left(j+1\right)\theta\right)}\right]\sin{\left(\left(t-j-k-\cfrac{3}{2}\right)\theta\right)}.
\end{eqnarray}
Here, we use the following theorem, which is a generalization of a well-known equality derived from the Dirichlet kernel:
\begin{theorem}[\cite{K18}]
For any natural number $t$ and real numbers $a$ and $\theta\not\equiv 0\ (mod\ 2\pi)$,
\begin{eqnarray}
\sum_{j=0}^{t-1}\sin{(a+j\theta)}=\cfrac{\sin{\left(\cfrac{t}{2}\theta\right)}\sin{\left(a+\cfrac{t-1}{2}\theta\right)}}{\sin{\left(\cfrac{\theta}{2}\right)}}.
\end{eqnarray}
\end{theorem}
\noindent
From Eq.~(\ref{amp1}),
\begin{eqnarray}
\sum_{x\in\mathcal{X}_{1,0}}P_{\rm suc}^{(x)}(\vec{\epsilon},t)=\left|\langle 1'_L|[D\tilde{O}_f(\tilde{\epsilon})]^t|\psi(0)\rangle\right|^2&\simeq&\cfrac{\left|\sin{\left(\left(t+\cfrac{1}{2}\right)\theta\right)}+\cfrac{\pi^2}{4}\left(s_1+s_2\right)\tilde{\epsilon}^2-i\cfrac{\pi}{2}s_1\tilde{\epsilon}\right|^2}{2}\\
\label{prob1'}
&\simeq&\cfrac{\sin^2{\left(\left(t+\cfrac{1}{2}\right)\theta\right)}+\cfrac{\pi^2}{4}\left[2\left(s_1+s_2\right)\sin{\left(\left(t+\cfrac{1}{2}\right)\theta\right)}+s_1^2\right]\tilde{\epsilon}^2}{2}.\ \ \ \ \
\end{eqnarray}
In a similar manner,
\begin{eqnarray}
\nonumber
&&\langle 1''_L|[D\tilde{O}_f(\tilde{\epsilon})]^t|\psi(0)\rangle\\
\nonumber
&\simeq&\cfrac{1}{\sqrt{2}}\begin{pmatrix}
0 & 1 & -1 \\
\end{pmatrix}
\left[\begin{pmatrix}
\cos{\left(\left(t+\cfrac{1}{2}\right)\theta\right)} \\
\sin{\left(\left(t+\cfrac{1}{2}\right)\theta\right)} \\
0 \\
\end{pmatrix}\right.\\
\nonumber
&&+\left(-i\cfrac{\pi}{2}\tilde{\epsilon}+\cfrac{\pi^2}{4}\tilde{\epsilon}^2\right)\sum_{j=0}^{t-1}\begin{pmatrix}
0 & \sin{\left(\left(j+1\right)\theta\right)} & -\sin{\left(\left(j+1\right)\theta\right)} \\
0 & -\cos{\left(\left(j+1\right)\theta\right)} & \cos{\left(\left(j+1\right)\theta\right)} \\
0 & 1 & -1 \\
\end{pmatrix}
\begin{pmatrix}
\cos{\left(\left(t-j-\cfrac{1}{2}\right)\theta\right)} \\
\sin{\left(\left(t-j-\cfrac{1}{2}\right)\theta\right)} \\
0 \\
\end{pmatrix}\\
&&+\cfrac{\pi^2}{4}\tilde{\epsilon}^2\sum_{j=0}^{t-2}\sum_{k=0}^{t-j-2}\left[1+\cos{\left((k+1)\theta\right)}\right]\left.\begin{pmatrix}
0 & \sin{\left(\left(j+1\right)\theta\right)} & -\sin{\left(\left(j+1\right)\theta\right)} \\
0 & -\cos{\left(\left(j+1\right)\theta\right)} & \cos{\left(\left(j+1\right)\theta\right)} \\
0 & 1 & -1 \\
\end{pmatrix}
\begin{pmatrix}
\cos{\left(\left(t-j-k-\cfrac{3}{2}\right)\theta\right)} \\
\sin{\left(\left(t-j-k-\cfrac{3}{2}\right)\theta\right)} \\
0 \\
\end{pmatrix}\right]\ \ \ \ \ \\
\nonumber
&=&\cfrac{\sin{\left(\left(t+\cfrac{1}{2}\right)\theta\right)}}{\sqrt{2}}-\left(-i\cfrac{\pi}{2}\tilde{\epsilon}+\cfrac{\pi^2}{4}\tilde{\epsilon}^2\right)\sum_{j=0}^{t-1}\cfrac{1+\cos{\left(\left(j+1\right)\theta\right)}}{\sqrt{2}}\sin{\left(\left(t-j-\cfrac{1}{2}\right)\theta\right)}\\
&&-\cfrac{\pi^2}{4}\tilde{\epsilon}^2\sum_{j=0}^{t-2}\sum_{k=0}^{t-j-2}\left[1+\cos{\left(\left(k+1\right)\theta\right)}\right]\cfrac{1+\cos{\left(\left(j+1\right)\theta\right)}}{\sqrt{2}}\sin{\left(\left(t-j-k-\cfrac{3}{2}\right)\theta\right)},
\end{eqnarray}
and thus
\begin{eqnarray}
\sum_{x\in\mathcal{X}_{1,\tilde{\epsilon}}}P_{\rm suc}^{(x)}(\vec{\epsilon},t)=\left|\langle 1''_L|[D\tilde{O}_f(\tilde{\epsilon})]^t|\psi(0)\rangle\right|^2&\simeq&\cfrac{\left|\sin{\left(\left(t+\cfrac{1}{2}\right)\theta\right)}-\cfrac{\pi^2}{4}\left(s_3+s_4\right)\tilde{\epsilon}^2+i\cfrac{\pi}{2}s_3\tilde{\epsilon}\right|^2}{2}\\
\label{prob1''}
&\simeq&\cfrac{\sin^2{\left(\left(t+\cfrac{1}{2}\right)\theta\right)}+\cfrac{\pi^2}{4}\left[-2\left(s_3+s_4\right)\sin{\left(\left(t+\cfrac{1}{2}\right)\theta\right)}+s_3^2\right]\tilde{\epsilon}^2}{2},\ \ \ \ \ \ \
\end{eqnarray}
where
\begin{eqnarray}
s_3\equiv\sum_{j=0}^{t-1}\left[1+\cos{\left(\left(j+1\right)\theta\right)}\right]\sin{\left(\left(t-j-\cfrac{1}{2}\right)\theta\right)}=\cfrac{\sin^2{\left(\cfrac{t\theta}{2}\right)}}{\sin{\left(\cfrac{\theta}{2}\right)}}+\cfrac{t}{2}\sin{\left(\left(t+\cfrac{1}{2}\right)\theta\right)}-\cfrac{\sin{\left(t\theta\right)}\sin{\left(\cfrac{\theta}{2}\right)}}{2\sin{\left(\theta\right)}}
\end{eqnarray}
and
\begin{eqnarray}
s_4\equiv\sum_{j=0}^{t-2}\sum_{k=0}^{t-j-2}\left[1+\cos{\left(\left(k+1\right)\theta\right)}\right]\left[1+\cos{\left(\left(j+1\right)\theta\right)}\right]\sin{\left(\left(t-j-k-\cfrac{3}{2}\right)\theta\right)}.
\end{eqnarray}

To evaluate how largely the marked items in $\mathcal{X}_{1,0}$ are prioritized than the other marked items in $\mathcal{X}_{1,\tilde{\epsilon}}$, it would be informative to calculate the gap $\sum_{x\in\mathcal{X}_{1,0}}P_{\rm suc}^{(x)}(\vec{\epsilon},t)-\sum_{x\in\mathcal{X}_{1,\tilde{\epsilon}}}P_{\rm suc}^{(x)}(\vec{\epsilon},t)$.
From Eqs.~(\ref{prob1'}) and (\ref{prob1''}),
\begin{eqnarray}
\nonumber
&&\sum_{x\in\mathcal{X}_{1,0}}P_{\rm suc}^{(x)}(\vec{\epsilon},t)-\sum_{x\in\mathcal{X}_{1,\tilde{\epsilon}}}P_{\rm suc}^{(x)}(\vec{\epsilon},t)\\
&=&\left|\langle 1'_L|[D\tilde{O}_f(\tilde{\epsilon})]^t|\psi(0)\rangle\right|^2-
\left|\langle 1''_L|[D\tilde{O}_f(\tilde{\epsilon})]^t|\psi(0)\rangle\right|^2\\
&\simeq&\cfrac{\pi^2}{4}\left[(s_1+s_3+s_2+s_4)\sin{\left(\left(t+\cfrac{1}{2}\right)\theta\right)}+\cfrac{(s_1+s_3)(s_1-s_3)}{2}\right]\tilde{\epsilon}^2\\
\nonumber
&=&\cfrac{\pi^2}{4}\left\{\left[2\cfrac{\sin^2{\left(\cfrac{t\theta}{2}\right)}}{\sin{\left(\cfrac{\theta}{2}\right)}}+\left(s_2+s_4\right)\right]\sin{\left(\left(t+\cfrac{1}{2}\right)\theta\right)}+\cfrac{2\cfrac{\sin^2{\left(\cfrac{t\theta}{2}\right)}}{\sin{\left(\cfrac{\theta}{2}\right)}}\left[-t\sin{\left(\left(t+\cfrac{1}{2}\right)\theta\right)}+\cfrac{\sin{\left(t\theta\right)}\sin{\left(\cfrac{\theta}{2}\right)}}{\sin{\left(\theta\right)}}\right]}{2}\right\}\tilde{\epsilon}^2\\
&&\\
&=&\cfrac{\pi^2}{4}\left\{\left[-(t-2)\cfrac{\sin^2{\left(\cfrac{t\theta}{2}\right)}}{\sin{\left(\cfrac{\theta}{2}\right)}}+\left(s_2+s_4\right)\right]\sin{\left(\left(t+\cfrac{1}{2}\right)\theta\right)}+\cfrac{\sin^2{\left(\cfrac{t\theta}{2}\right)}\sin{\left(t\theta\right)}}{\sin{\left(\theta\right)}}\right\}\tilde{\epsilon}^2\\
\label{Bfin}
&=&\cfrac{\pi^2}{4}\left[\cfrac{t\sin{\left(\theta\right)}-\sin{\left(t\theta\right)}\cos{\left(\theta\right)}}{2\sin{\left(\theta\right)}\sin{\left(\cfrac{\theta}{2}\right)}}\sin{\left(\left(t+\cfrac{1}{2}\right)\theta\right)}+\cfrac{\sin^2{\left(\cfrac{t\theta}{2}\right)}\sin{\left(t\theta\right)}}{\sin{\left(\theta\right)}}\right]\tilde{\epsilon}^2,
\end{eqnarray}
where we have used
\begin{eqnarray}
s_2+s_4&=&2\sum_{j=0}^{t-2}\sum_{k=0}^{t-j-2}\left[1+\cos{\left(\left(k+1\right)\theta\right)}\right]\sin{\left(\left(t-j-k-\cfrac{3}{2}\right)\theta\right)}\\
&=&2\sum_{j=0}^{t-2}\sum_{k=0}^{t-j-2}\left[1+\cos{\left(\left(j+1\right)\theta\right)}\right]\sin{\left(\left(t-j-k-\cfrac{3}{2}\right)\theta\right)}\\
&=&2\sum_{j=0}^{t-2}\left[1+\cos{\left(\left(j+1\right)\theta\right)}\right]\cfrac{\sin^2{\left(\cfrac{t-j-1}{2}\theta\right)}}{\sin{\left(\cfrac{\theta}{2}\right)}}\\
&=&\cfrac{4}{\sin{\left(\cfrac{\theta}{2}\right)}}\sum_{j=0}^{t-2}\cos^2{\left(\cfrac{j+1}{2}\theta\right)}\sin^2{\left(\cfrac{t-j-1}{2}\theta\right)}\\
&=&\cfrac{1}{\sin{\left(\cfrac{\theta}{2}\right)}}\sum_{j=0}^{t-2}\left[\sin{\left(\cfrac{t\theta}{2}\right)}+\sin{\left(\cfrac{t-2(j+1)}{2}\theta\right)}\right]^2\\
&=&\cfrac{(t-1)\sin^2{\left(\cfrac{t\theta}{2}\right)}+2\sin{\left(\cfrac{t\theta}{2}\right)}\left[\sum_{j=0}^{t-2}\sin{\left(\cfrac{t-2(j+1)}{2}\theta\right)}\right]+\left[\sum_{j=0}^{t-2}\cfrac{1-\cos{\left(\left(t-2j-2\right)\theta\right)}}{2}\right]}{\sin{\left(\cfrac{\theta}{2}\right)}}\\
&=&\cfrac{(t-1)\sin^2{\left(\cfrac{t\theta}{2}\right)}+\left[\cfrac{t-1}{2}-\cfrac{\sin{\left(\left(t-1\right)\theta\right)}}{2\sin{\left(\theta\right)}}\right]}{\sin{\left(\cfrac{\theta}{2}\right)}}\\
&=&(t-2)\cfrac{\sin^2{\left(\cfrac{t\theta}{2}\right)}}{\sin{\left(\cfrac{\theta}{2}\right)}}+\cfrac{t\sin{\left(\theta\right)}-\sin{\left(t\theta\right)}\cos{\left(\theta\right)}}{2\sin{\left(\theta\right)}\sin{\left(\cfrac{\theta}{2}\right)}}
\end{eqnarray}
to derive Eq.~(\ref{Bfin}).
Since $0<\sin{(\theta/2)}\le\cos{(\theta/2)}$ from $1\le m\le n/2$, Eq.~(\ref{Bfin}) implies that $\sum_{x\in\mathcal{X}_{1,0}}P_{\rm suc}^{(x)}(\vec{\epsilon},t)>\sum_{x\in\mathcal{X}_{1,\tilde{\epsilon}}}P_{\rm suc}^{(x)}(\vec{\epsilon},t)$ holds when $\tilde{\epsilon}\neq 0$ by setting $t$ so that (i) ${\rm min}\left\{\sin{\left(t\theta\right)},\cos{\left(t\theta\right)}\right\}\ge0$ and (ii) $t>1/\tan{(\theta)}$.

\section{REASON FOR SELECTING $\rho_0$}
\label{C}
The purpose in Sec.~\ref{IIIC} is to examine the effect of the coherence in the input state.
Therefore, a quantum state that is almost the same as the coherent state $|\psi(0)\rangle$ except for the coherence would be proper as an incoherent state to be compared.
From Eq.~(\ref{ini2}), we notice that the two marked items $\tilde{x}_1$ and $\tilde{x}_2$ are treated equally in $|\psi(0)\rangle$.
Furthermore, all the probability amplitudes are real, and the probability of incorrect answers (i.e., $|0_L\rangle$) being observed is written as a non-negative integer over the database size $1000$.
To mimic these properties of $|\psi(0)\rangle$ in an incoherent state, we assume that $\rho_0$ is given as
\begin{eqnarray}
\nonumber
&&\cfrac{1}{2}\left(\sqrt{\cfrac{a}{1000}}|0_L\rangle+\sqrt{1-\cfrac{a}{1000}}|\tilde{x}_1\rangle\right)\left(\sqrt{\cfrac{a}{1000}}\langle0_L|+\sqrt{1-\cfrac{a}{1000}}\langle\tilde{x}_1|\right)\\
\label{incoherentcand}
&+&\cfrac{1}{2}\left(\sqrt{\cfrac{a}{1000}}|0_L\rangle+\sqrt{1-\cfrac{a}{1000}}|\tilde{x}_2\rangle\right)\left(\sqrt{\cfrac{a}{1000}}\langle0_L|+\sqrt{1-\cfrac{a}{1000}}\langle\tilde{x}_2|\right)
\end{eqnarray}
with $0\le a\le 1000$ being a non-negative integer.

We show that $\rho_0$ in Eq.~(\ref{rho0}) is the quantum state closest to $|\psi(0)\rangle$ under this assumption.
The fidelity between the quantum state in Eq.~(\ref{incoherentcand}) and $|\psi(0)\rangle$ in Eq.~(\ref{ini2}) is
\begin{eqnarray}
\label{fidelityC}
\left[\sqrt{\cfrac{a}{1000}\times\cfrac{499}{500}}+\sqrt{\cfrac{1}{1000}\left(1-\cfrac{a}{1000}\right)}\right]^2=\cfrac{\left(\sqrt{998a}+\sqrt{1000-a}\right)^2}{10^6}.
\end{eqnarray}
Eq.~(\ref{fidelityC}) is maximized when $\sqrt{998a}+\sqrt{1000-a}$ is maximized, and hence the fidelity takes its maximum value $(997003+6\sqrt{110778})/10^6\simeq0.999$ at $a=999$.
In this case, the quantum state in Eq.~(\ref{incoherentcand}) becomes Eq.~(\ref{rho0}).


\begin{thebibliography}{99}
\bibitem{M09}M. Mosca, Quantum Algorithms, in {\it Encyclopedia of Complexity and Systems Science} (Springer, New York, 2009), p. 7088.
\bibitem{HHL09}A. W. Harrow, A. Hassidim, and S. Lloyd, Quantum Algorithm for Linear Systems of Equations, Phys. Rev. Lett. {\bf 103}, 150502 (2009).
\bibitem{MRTC21}J. M. Martyn, Z. M. Rossi, A. K. Tan, and I. L. Chuang, Grand Unification of Quantum Algorithms, PRX Quantum {\bf 2}, 040203 (2021).
\bibitem{CVZLL98}I. L. Chuang, L. M. K. Vandersypen, X. Zhou, D. W. Leung, and S. Lloyd, Experimental realization of a quantum algorithm, Nature (London) {\bf 393}, 143 (1998).
\bibitem{JMH98}J. A. Jones, M. Mosca, and R. H. Hansen, Implementation of a quantum search algorithm on a quantum computer, Nature (London) {\bf 393}, 344 (1998).
\bibitem{VSBYSC01}L. M. K. Vandersypen, M. Steffen, G. Breyta, C. S. Yannoni, M. H. Sherwood, and I. L. Chuang, Experimental realization of Shor's quantum factoring algorithm using nuclear magnetic resonance, Nature (London) {\bf 414}, 883 (2001).
\bibitem{WRRSWVAZ05}P. Walther, K. J. Resch, T. Rudolph, E. Schenck, H. Weinfurter, V. Vedral, M. Aspelmeyer, and A. Zeilinger, Experimental one-way quantum computing, Nature (London) {\bf 434}, 169 (2005).
\bibitem{ZSCXLGZXDHWYZLPWLZ17}Y. Zheng, C. Song, M.-C. Chen, B. Xia, W. Liu, Q. Guo, L. Zhang, D. Xu, H. Deng, K. Huang, Y. Wu, Z. Yan, D. Zheng, L. Lu, J.-W. Pan, H. Wang, C.-Y. Lu, and X. Zhu, Solving Systems of Linear Equations with a Superconducting Quantum Processor, Phys. Rev. Lett. {\bf 118}, 210504 (2017).
\bibitem{G97}L. K. Grover, Quantum Mechanics Helps in Searching for a Needle in a Haystack, Phys. Rev. Lett. {\bf 79}, 325 (1997).
\bibitem{BHT98}G. Brassard, P. H{\O}yer, and A. Tapp, Quantum cryptanalysis of hash and claw-free functions, in {\it Proc. of the 3rd Latin American Symposium} (Springer, Campinas, 1998), p. 163.
\bibitem{WNHT20}A. Y. Wei, P. Naik, A. W. Harrow, and J. Thaler, Quantum algorithms for jet clustering, Phys. Rev. D {\bf 101}, 094015 (2020).
\bibitem{DHLT21}Y. Du, M.-H. Hsieh, T. Liu, and D. Tao, A Grover-search based quantum learning scheme for classification, New J. Phys. {\bf 23}, 023020 (2021).
\bibitem{BBW05}W. P. Baritompa, D. W. Bulger, and G. R. Wood, Grover's Quantum Algorithm Applied to Global Optimization, SIAM Journal on Optimization {\bf 15}, 1170 (2005).
\bibitem{DHHM06}C. D\"{u}rr, M. Heiligman, P. HOyer, and M. Mhalla, Quantum Query Complexity of Some Graph Problems, SIAM J. Comput. {\bf 35}, 1310 (2006).
\bibitem{G96}L. K. Grover, A fast quantum mechanical algorithm for database search, in {\it Proc. of the 28th Annual Symposium on Theory of Computing} (ACM, Philadelphia, 1996), p. 212.
\bibitem{BBBV97}C. H. Bennett, E. Bernstein, G. Brassard, and U. Vazirani, Strengths and Weaknesses of Quantum Computing, SIAM J. Comput. {\bf 26}, 1510 (1997).
\bibitem{BBHT99}M. Boyer, G. Brassard, P. H{\o}yer, and A. Tapp, Tight Bounds on Quantum Searching, Fortschr. Phys. {\bf 46}, 493 (1998).
\bibitem{Z99}C. Zalka, Grover's quantum searching algorithm is optimal, Phys. Rev. A {\bf 60}, 2746 (1999).
\bibitem{H00}P. H{\o}yer, Arbitrary phases in quantum amplitude amplification, Phys. Rev. A {\bf 62}, 052304 (2000).
\bibitem{L01}G. L. Long, Grover algorithm with zero theoretical failure rate, Phys. Rev. A {\bf 64}, 022307 (2001).
\bibitem{BHMT02}G. Brassard, P. Hoyer, M. Mosca, and A. Tapp, Quantum amplitude amplification and estimation, Contemp. Math. {\bf 305}, 53 (2002).
\bibitem{RJS22}T. Roy, L. Jiang, and D. I. Schuster, Deterministic Grover search with a restricted oracle, Phys. Rev. Research {\bf 4}, L022013 (2022).
\bibitem{RJS23}T. Roy, L. Jiang, and D. I. Schuster, Erratum: Deterministic Grover search with a restricted oracle, Phys. Rev. Research {\bf 5}, 029002 (2023).
\bibitem{BCWZ99}H. Buhrman, R. Cleve, R. De Wolf, and C. Zalka, Bounds for small-error and zero-error quantum algorithms, in {\it Proc. of the 40th Annual Symposium on Foundations of Computer Science} (IEEE, New York, 1999), p. 358.
\bibitem{PS08}L. Panchi and L. Shiyong, Grover quantum searching algorithm based on weighted targets, Journal of Systems Engineering and Electronics {\bf 19}, 363 (2008).
\bibitem{RK23}K. Roy and M.-K. Kim, Applying Quantum Search Algorithm to Select Energy-Efficient Cluster Heads in Wireless Sensor Networks, Electronics {\bf 12}, 63 (2023).
\bibitem{RS08}O. Regev and L. Schiff, Impossibility of a Quantum Speed-Up with a Faulty Oracle, in {\it Proc. of the 35th International Colloquium on Automata, Languages and Programming} (Springer, Reykjavik, 2008), p. 773.
\bibitem{R23}A. Rosmanis, Quantum Search with Noisy Oracle, arXiv:2309.14944.
\bibitem{HMW03}P. H{\o}yer, M. Mosca, and R. de Wolf, Quantum Search on Bounded-Error Inputs, in {\it Proc. of the 30th International Colloquium on Automata, Languages and Programming} (Springer, Eindhoven, 2003), p. 291.
\bibitem{LLZT20}G. L. Long, Y. S. Li, W. L. Zhang, and C. C. Tu, Dominant gate imperfection in Grover's quantum search algorithm, Phys. Rev. A {\bf 61}, 042305 (2000).
\bibitem{SBW03}N. Shenvi, K. R. Brown, and K. B. Whaley, Effects of a random noisy oracle on search algorithm complexity, Phys. Rev. A {\bf 68}, 052313 (2003).
\bibitem{LSZN99}G. L. Long, Y. S. Li, W. L. Zhang, and L. Niu, Phase matching in quantum searching, Phys. Lett. A {\bf 262}, 27 (1999).
\bibitem{LL07}P. Li and S. Li, Phase matching in Grover's algorithm, Phys. Lett. A {\bf 366}, 42 (2007).
\bibitem{TDNTK08}F. M. Toyama, W. van Dijk, Y. Nogami, M. Tabuchi, and Y. Kimura, Multiphase matching in the Grover algorithm, Phys. Rev. A {\bf 77}, 042324 (2008).
\bibitem{TD24}H. Tonchev and P. Danev, Robustness of different modifications of Grover's algorithm based on generalized Householder reflections with different phases, Results Phys. {\bf 59}, 107595 (2024).
\bibitem{ABNR12}A. Ambainis, A. Ba\v{c}kurs, N. Nahimovs, and A. Rivosh, Grover's Algorithm with Errors, in {\it Proc. of the 8th Doctoral Workshop on Mathematical and Engineering Methods in Computer Science} (Springer, Znojmo, 2012), p. 180.
\bibitem{KNR18}D. Kravchenko, N. Nahimovs, and A. Rivosh, Grover's Search with Faults on Some Marked Elements, International Journal of Foundations of Computer Science {\bf 29}, 647 (2018).
\bibitem{DZKK24}S. Dowarah, C. Zhang, V. Khemani, and M. Kolodrubetz, Phases and phase transition in Grover's algorithm with systematic noise, Phys. Rev. A {\bf 111}, 042603 (2025).
\bibitem{SKH22}Y. Seo, Y. Kang, and J. Heo, Quantum Search Algorithm for Weighted Solutions, IEEE Access {\bf 10}, 16209 (2022).
\bibitem{SW24}Y. Sun and L.-A. Wu, Quantum search algorithm on weighted databases, Sci. Rep. {\bf 14}, 30169 (2024).
\bibitem{LMP03}C. Lavor, L.R.U. Manssur, and R. Portugal, Grover's Algorithm: Quantum Database Search, arXiv:quant-ph/0301079.
\bibitem{GK17}P. R. Giri and V. E. Korepin, A review on quantum search algorithms, Quantum Inf. Process. {\bf 16}, 315 (2017).
\bibitem{BCP14}T. Baumgratz, M. Cramer, and M. B. Plenio, Quantifying Coherence, Phys. Rev. Lett. {\bf 113}, 140401 (2014).
\bibitem{NC10}M. A. Nielsen and I. L. Chuang, {\it Quantum Computation and Quantum Information 10th Anniversary Edition} (Cambridge University Press, Cambridge, 2010).
\bibitem{SBCJKMMPBS24}Z. Sun, G. Boyd, Z. Cai, H. Jnane, B. Koczor, R. Meister, R. Minko, B. Pring, S. C. Benjamin, and N. Stamatopoulos, Low-depth phase oracle using a parallel piecewise circuit, Phys. Rev. A {\bf 111}, 062420 (2025).
\bibitem{YLC14}T. J. Yoder, G. H. Low, and I. L. Chuang, Fixed-Point Quantum Search with an Optimal Number of Queries, Phys. Rev. Lett. {\bf 113}, 210501 (2014).
\bibitem{CBD23}S. R. Chowdhury, S. Baruah, and B. Dikshit, Phase matching in quantum search algorithm, EPL {\bf 141}, 58001 (2023).
\bibitem{PR99}B. Pablo-Norman and M. Ruiz-Altaba, Noise in Grover's quantum search algorithm, Phys. Rev. A {\bf 61}, 012301 (1999).
\bibitem{SMB03}D. Shapira, S. Mozes, and O. Biham, Effect of unitary noise on Grover's quantum search algorithm, Phys. Rev. A {\bf 67}, 042301 (2003).
\bibitem{LYW23}J. Leng, F. Yang, and X.-B. Wang, Improving D2p Grover's algorithm to reach performance upper bound under phase noise, Phys. Rev. Research {\bf 5}, 023202 (2023).
\bibitem{K18}M. P. Knapp, Sines and Cosines of Angles in Arithmetic Progression, Mathematics Magazine {\bf 82}, 371 (2018).
\end{thebibliography}
\end{document}